\documentclass[11pt,onecolumn,conference]{IEEEtran}
\usepackage{graphicx}
\usepackage{amssymb}
\usepackage{amsmath}
\usepackage{amssymb}
\usepackage{array}
\usepackage{xy}
\usepackage{multirow}
\usepackage{graphicx}
\usepackage{amssymb}
\usepackage{amsmath}
\usepackage{amssymb}
\usepackage{array}
\usepackage{xy}
\usepackage{multirow}
\usepackage{authblk}
\usepackage{authblk}
\usepackage{mathptmx}

\newcommand\blfootnote{\xdef\@thefnmark{}\@footnotetext}

\title{\textsl{On Optimal and Fair Service Allocation in Mobile Cloud Computing}}
\author[1]{M. Reza Rahimi}
\author[1]{Nalini Venkatasubramanian}
\author[1]{Sharad Mehrotra}
\author[2]{Athanasios V. Vasilakos}
\affil[1]{School of Information and Computer Science, University of California, Irvine, USA.}
\affil[2]{National Technical University of Athens, Athens, Greece.}
\makeatletter
\renewcommand\AB@affilsepx{, \protect\Affilfont}
\makeatother
\affil[1]{{\{mrrahimi, nalini, sharad\}@ics.uci.edu}} \affil[2]{{vasilako@ath.forthnet.gr}}

\begin{document}
\maketitle
\vspace{-.15in}
\begin{abstract}
This paper studies the optimal and fair service allocation for a variety of mobile applications (single or group and collaborative mobile applications) in mobile cloud computing. We exploit the observation that using tiered clouds, i.e. clouds at multiple levels (local and public) can increase the performance and scalability of mobile applications. We proposed a novel framework to model mobile applications as a \textsl{location-time workflows} (LTW) of tasks; here users mobility patterns are translated to mobile service usage patterns. We show that an optimal mapping of LTWs to tiered cloud resources considering multiple QoS goals such application delay, device power consumption and user cost/price is an NP-hard problem for both single and group-based applications. We propose an efficient heuristic algorithm called \textsl{MuSIC} that is able to perform well (73\% of optimal, 30\% better than simple strategies), and scale well to a large number of users while ensuring high mobile application QoS. We evaluate MuSIC and the 2-tier mobile cloud approach via implementation (on real world clouds) and extensive simulations using rich mobile applications like intensive signal processing, video streaming and multimedia file sharing applications. Our experimental and simulation results indicate that MuSIC supports scalable operation (100+ concurrent users executing complex workflows) while improving QoS. We observe about 25\% lower delays and power (under fixed price constraints) and about 35\% decrease in price (considering fixed delay) in comparison to only using the public cloud. Our studies also show that MuSIC performs quite well under different mobility patterns, e.g. random waypoint and Manhattan models.
\end{abstract}

\section{Introduction} \label{introduction}

The rapid explosion in demand for rich mobile applications  has created the need for new platforms and architectures that can cope with the scalability and QoS needs of a growing mobile user population. One of the main bottlenecks in ensuring mobile QoS is the level of wireless connectivity offered by last hop access networks such as 3G and Wi-Fi. These networks exhibit varying characteristics. For example, 3G networks offer wide area ubiquitous connectivity; however, 3G connections are known to suffer from \emph{long delay} and \emph{slow data transfers} \cite{MAUI_2010}, \cite{Sokol_2013}, \cite{MAPCloud_2012}, \cite{Music_2013} resulting in increased power consumption and cost at the user side.  In contrast, Wi-Fi  deployments, e.g. 802.11 hotspots, exhibit low communication latencies/delays; devices connected to or collocated with Wi-Fi access points can be used to form a nearby local cloud \cite{CloneCloud_2011}, \cite{MAUI_2010}, \cite{cloudstorm_2012}. Using local only solutions with Wi-Fi networks creates \emph{scalability} and \emph{access} issues as the number of users increases. The second key issue is that rich mobile applications often require significant storage and  processing  abilities (e.g. content transcoding, caching, data interpretation) - despite advances in device technology, resources (energy, storage, processing) at the mobile host are limited.

\textbf{M}obile \textbf{C}loud \textbf{C}omputing (MCC) platforms aim to overcome the resource limitations of mobile devices and networks by leveraging resources available in distributed cloud environments. The goal is to offload  compute and data intensive tasks from resource-poor mobile devices to cloud nodes. Recent market studies (e.g from  Juniper Research \cite{Juniper_2010}) indicate that the market for cloud-based mobile applications will grow 88\% from 400 million in 2009 to 9.5 billion in 2014. A similar forecast made by ABI \cite{ABI_2009}, predicts that the number of MCC subscribers worldwide is expected to grow rapidly over the next five years, rising from 42.8 million subscribers in 2008 to over 998 million in 2014.

In our prior work \cite{MAPCloud_2012}, \cite{RezaWoWMoM_2012}, \cite{Music_2013} we discuss the role of public and local clouds in enabling scalable MCC.  While public clouds provide resources at scale; there is a limited number of public cloud data centers within \emph{close proximity} of mobile users resulting in large communication latencies. Recent efforts \cite{Satyanarayanan_2009}, \cite{MAUI_2010}, \cite{CloneCloud_2011}, \cite{MAPCloud_2012}, \cite{Sokol_2013}, \cite{Music_2013}, \cite{cloudstorm_2012} have demonstrated the role of local resources within close proximity of the mobile user in ensuring improved application performance.

The mobility of users introduces new complexities in ensuring QoS in MCC applications. As the number and speed of mobile users increase, mobile applications are faced with increased  \emph{latencies} and reduced \emph{reliability}. As a user moves, the physical distance between the user and the cloud resources originally provisioned changes causing additional delays. Similarly, the lack of effective handoff mechanisms in WiFi networks as the user moves rapidly causes an increase in the number of \emph{packet losses} \cite{3gwifi_2010}, \cite{Satyanarayanan_2009}. In other words, user mobility, if not addressed properly, can  result in suboptimal resource mapping choices and ultimately in diminished application QoS.

In our pervious works \cite{MAPCloud_2012}, \cite{RezaWoWMoM_2012}, \cite{Music_2013}, we have developed MAPCloud middleware framework that synergistically combines the capabilities of local clouds and public in \emph{2-Tier architecture}, and users mobility patterns. In this paper we extend our work to support \emph{group-based or collaborative mobile applications} considering their mobility patterns. In this class of mobile applications users are using shared services (like social network type of application and shared storage) and our goal is to optimally assign services to have higher \emph{group utility}.

\textbf{Key Contributions :} In this paper, we focus on developing efficient techniques and algorithms for dynamic mapping of services for single and group-based mobile applications. We aim to meet the multidimensional QoS needs of mobile users. The main contributions of this paper are as follow:
\begin{enumerate}
\item We extend our notion of a \emph{location-time workflow} (LTW) as the modeling framework in our previous work \cite{Music_2013} to model mobile applications and capture user mobility for group-based application. Within this framework, we formally model mobile service usage patterns as a function of location and time. Based on this modeling framework we formally model optimal service allocation problems for two different classes of mobile applications one is single user and the other one is group-based collaborative mobile applications (\ref{MathModelling}).
\item Given a mobile application execution expressed as a LTW, we optimally partition the execution of the  location-time workflow in the 2-tier architecture based on a \emph{utility metrics} (for both single and collaborative mobile applications) that combines \emph{service price, power consumption and delay} of the mobile applications. We show that the resulting service allocation problems are NP-Hard and propose an efficient heuristic called \textbf{MuSIC} (\textbf{M}obility-Aware \textbf{S}ervice Allocat\textbf{I}on on \textbf{C}loud) to achieve a near optimal solution (Section \ref{ProposedAlgorithm}).
\item We extend our prototype of the system using \emph{Amazon WS} as the public cloud, a local campus cloud and Android devices. We implement 3 real world rich media mobile applications (mobile video streaming,  image and speech processing application, and multimedia file sharing. We use multimedia file sharing as an example for group-based mobile application). We evaluate our system under varying user mobility patterns including the \emph{random waypoint} and \emph{manhattan} models \cite{Mobility_2002}, \cite{RWP_2004}; our simulation and experimental results indicate that MuSIC scales to a large number of users and performs in average within 73\% of the optimal solution for both single and collaborative mobile applications. Our experiments indicate that MuSIC is tolerant to errors and uncertainty in predicting mobile user location-time workflows - we achieve 62\% in average for both single and group-based applications of optimal performance when the uncertainty of location-time workflow prediction is as high as 30\% (Section \ref{Evaluation}).
\item We also evaluate the performance of 2-tier cloud architecture under significant mobility in comparison to using the public cloud alone. Our results indicate that the 2-tier cloud architecture \emph{decreases} power consumption and delay in 20\% on average when the price is fixed and decreases the average user's price about 32\% (fixed value for delay and power consumption) in comparison to using only a public cloud for both single and group-based applications (section \ref{Evaluation}).
\end{enumerate}
We conclude with related efforts (Section \ref{relatedwork}) and future directions( Section \ref{Conclusion}).

\section{Modeling Service Allocation on the Tiered Cloud} \label{MathModelling}

\begin{figure}
\begin{center}
\includegraphics[width=3in,height=3in]{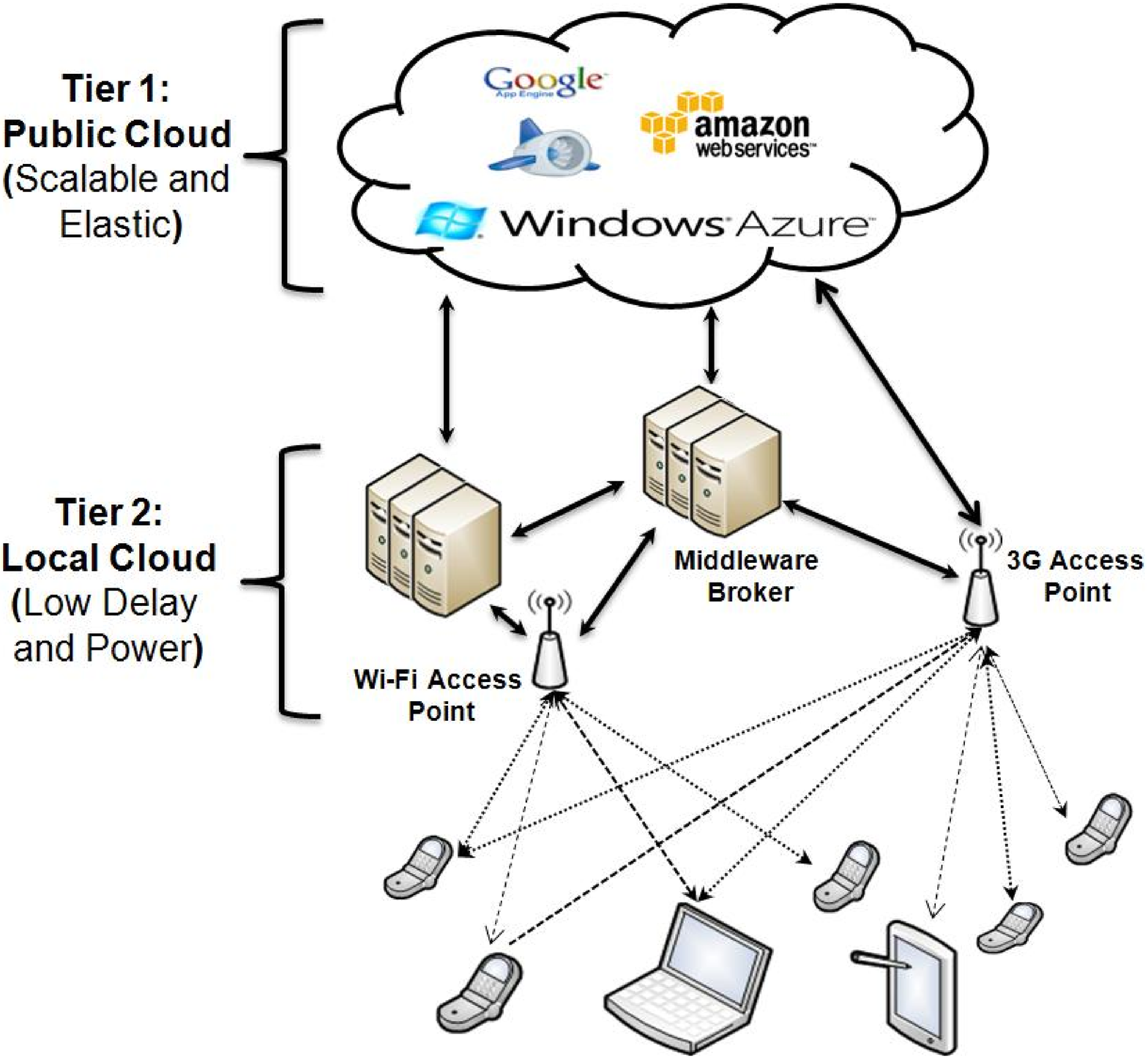}
\caption{2-Tier Mobile Cloud Architecture.}
\label{fig:2tiercloud}
\end{center}
\vspace{-.1in}
\end{figure}

Fig. \ref{fig:2tiercloud} shows the 2-tier cloud architecture \cite{MAPCloud_2012}, \cite{RezaWoWMoM_2012} for mobile applications . Tier 1 nodes in the system architecture represents public cloud services such as Amazon Web Services \cite{Amazon_WS},  Microsoft Azure \cite{Azure} and Google AppEngine \cite{Google_App_Engine}. Services provided by these vendors are highly  \emph{scalable} and \emph{available}; what they lack is the  ability to provide the \emph{fine grain location granularity} required for high performance mobile applications. This feature is provided by the second tier local cloud, that  consists of nodes that are connected to access points. Location information of these services are available at finer levels of granularity (campus and street level). Mobile users are typically connected to these local clouds through WiFi (via access points) or cellular (via 3G cell towers) connectivity -  our aim to to intelligently select which local and which public cloud resources to utilize for task offloading. In the following subsections, we develop concepts borrowed from service-oriented computing (SOC) literature \cite{Mabrouk_2009}, \cite{Zeng_2004}, \cite{Qos_2009}, \cite{Ou_2007} to formally define the notion of location-time workflows (LTW) for mobile applications and use the LTW concept to formulate the MCC service allocation problem.

\subsection{Mobile Application Modeling}

In this section we model cloud services, mobile users and mobile applications. Let's start by defining the concept of cloud service set.

\textsl{Def. 1}: \textbf{Cloud Service Set}: The set of all services ( e.g. compute, storage and software capabilities like multimedia streaming services, content transcoding services, etc ) provided by local and public cloud providers. $C_s$ is formally expressed as:

\vspace{-.1in}
\small
$$ C_s \triangleq  \{s_1,s_2,...,s_{|C_s|} \}$$
\normalsize

\textsl{Def. 2}: \textbf{Local Cloud Capacity}: Local cloud services can only accept a limited number of mobile client requests. We define a function $Cap(LC)$ which returns the  \textit{maximum} number of mobile clients that could be served using local cloud ($LC$).

\textsl{Def. 3}: \textbf{Location Map}:  is a partition of the $2-D$ space/region in which mobile hosts and cloud resources are located.  Given a $2D$ region in $R^2$, the location map $L$ is more formally defined as:

\vspace{-.15in}
\small
\begin{align}
\nonumber
L \triangleq \{l_1, l_2,...,l_{|L|}\},~\forall i,j \in \{1,2,...,|L|\},~l_i, l_j \subseteq \mathbb{R}^{2}\\
\nonumber
l_i \bigcap l_j=\emptyset,~\bigcup_{i=1}^{i=|L|}l_i=\mathbb{R}^{2}~~~~~~~~~~~~~~~~
\end{align}
\normalsize

We assign a vector to the center of location, depicted as ($l_i \Leftrightarrow \overrightarrow{l_i}$).

\textsl{Def. 4}: \textbf{User Service Set}: The set of all services that a user $u_k$ has on his own device (e.g. decoders, image editors etc.). It is represented as:

\vspace{-.15in}
\small
$$U_{k}^{s} \triangleq \{u_k^{s_1}, u_k^{s_2}, ..., u_k^{s_{|U_{k}^s |}}\}$$
\normalsize

\textsl{Def. 5}: \textbf{Mobile User Trajectory}: The trajectory of a mobile user, $u_k$, is represented as a list of tuples of the form $\{(l_k,\tau_1),...,(l_m, \tau_n)\}$ where $(l_j,\tau_i)$ implies that the mobile user is in location $l_j$ for time duration $\tau_i$.

\textsl{Def. 6}: \textbf{Center of Mobility}: $l^{u_k}_{cm}$ is the location where (or near where) a mobile user $u_k$ spends most of its time. It is calculated as follows:

\vspace{-.15in}
\small
\begin{align}
\nonumber
l^{u_k}_{cm} \in \{l_1, l_2,...,l_{|L|}\}~~~~~~~~~~~\\
\nonumber
l^{u_k}_{cm} \triangleq \min_{\overrightarrow{l_j} \in \{\overrightarrow{l_1}, \overrightarrow{l_2},..., \overrightarrow{l_{|L|}}\}} \|\frac{\sum_{i}\overrightarrow{l_i}\tau_i}{\sum_{i}\tau_i}-\overrightarrow{l_j}\|\\
\nonumber
\end{align}
\normalsize
\vspace{-.2in}

In this formula $\|~\|$ represents norm operation in $2D$ vector space.

\textsl{Def. 7}: \textbf{User Group Set}: It is defined as a group of users which have the \emph{shared services} for example in storage sharing and collaborative applications. It is presented by $G \subseteq \mathcal P \left({U}\right)$ in which $\mathcal P \left({U}\right)$ is the power set (the set that contains all subset of $U$, in which $U$ is the set of all users). We could define it formally as:

\vspace{-.1in}
\small
\begin{align}
\nonumber
G \triangleq \{ g_1,g_2,...,g_{|G|}\},~G \subseteq \mathcal P \left({U}\right)\\
\nonumber
\forall i \in \{1,2,...,|G|\},~g_i \subseteq U~~~~
\end{align}
\normalsize

\textsl{Def. 8}: \textbf{Center of Group Mobility}: $l^{g_i}_{cm}$ is the location that a mobile group users belong to $g_i$ spends most of their time in $l^{g_i}_{cm}$ or \emph{near} it. It could be formally defined as:

\vspace{-.1in}
\small
\begin{align}
\nonumber
l^{g_i}_{cm} \in \{l_1, l_2,...,l_{|L|}\}~~~
\nonumber
l^{g_i}_{cm} \triangleq \frac{1}{|g_i|}\sum_{u_k \in g_i} \overrightarrow{l_{cm}^{u_k}}
\nonumber
\end{align}
\normalsize
\vspace{-.1in}

In this case $|g_i|$ is the group size. Like center of mobility we can assign a \emph{vector} to the center of location of group. It could be shown as ($l^{g_i}_{cm} \Leftrightarrow \overrightarrow{l^{g_i}_{cm}} $).

\begin{figure}
\begin{center}
\includegraphics[width=3.25in,height=3.25in]{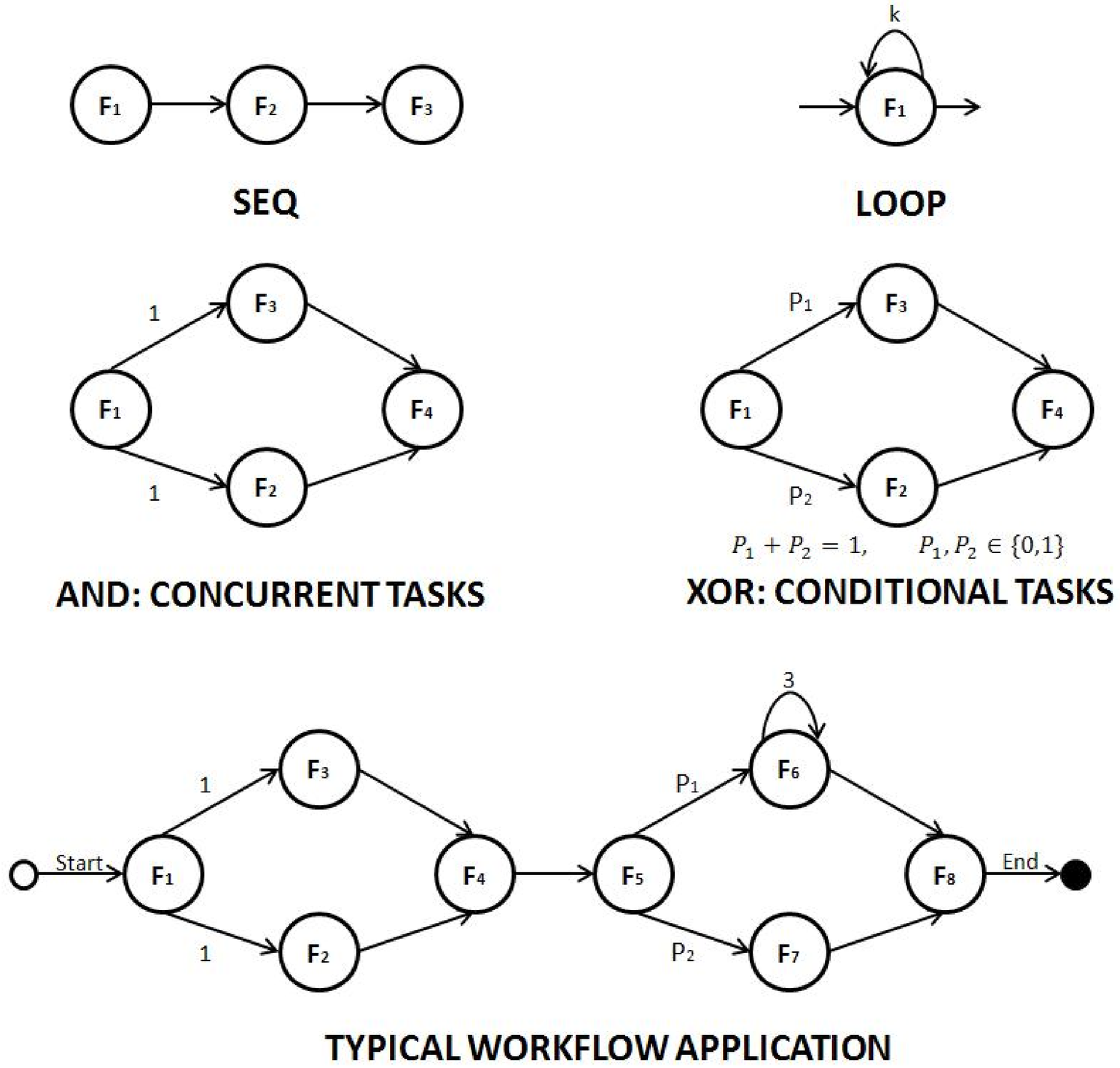}
\caption{ Different function patterns and sample workflow}
\label{fig:workflowpattern}
\end{center}
\vspace{-.1in}
\end{figure}

\textsl{Def. 9}: \textbf{Mobile Application Workflow}: A generic mobile application is modeled as a \emph{workflow} $w$ \cite{Zeng_2004}, \cite{MAPCloud_2012}, \cite{RezaWoWMoM_2012} consisting of a sequence of logical and precise steps, each of which is known as a \emph{Function}. A workflow begins at the start function and finishes in the final function. Functions in a workflow can be composed together in different patterns as shown in Fig. \ref{fig:workflowpattern}. The \textbf{SEQ} pattern indicates a sequential execution of the functions. The \textbf{AND} pattern models the parallel execution of the functions. \textbf{XOR} is a conditional execution of the functions and \textbf{LOOP} pattern indicates an iterative repetition of the functions. Each function is associated with a set of \emph{services} that are capable of realizing and implementing the function in the tiered cloud architecture.

For each Function $F_i$ in workflow $w$ we define $\chi_{F_i}$ as:  \small $$ \chi_{F_i}\triangleq \{s_k~|~s_k \in U^{s}\cup C_{s},~s_k~implements~F_i\}$$ \normalsize
Intuitively $\chi_{F_i}$ is the set of all services that could realize function $F_i$. For the workflow $w$ consisting of of $n$ tasks, the set $\Gamma$ describes all the feasible solutions or execution plans \cite{Zeng_2004}. It is defined as the cartesian product:

\small
$$ \Gamma_{w} \triangleq \chi_{F_1}\times \chi_{F_2} \times.....\times \chi_{F_n} $$
\normalsize

\textsl{Def. 10}: \textbf{Location-Time Workflow (LTW)}: We next combine the mobile application workflow concept above and with a user trajectory to model the mobile users and the requested services in their trajectory.

\begin{figure}[b]
\begin{center}
\includegraphics[width=3in,height=2in]{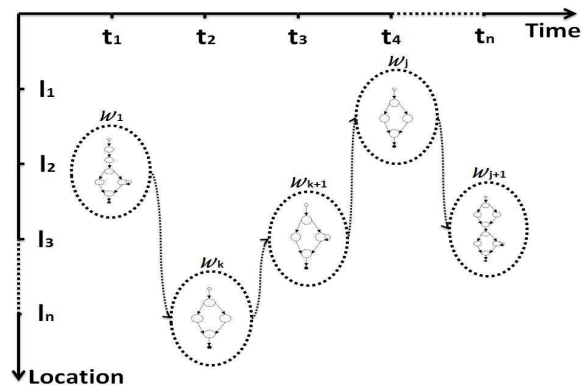}
\caption{ Location-Time Workflow}
\label{fig:spacetime}
\end{center}
\vspace{-.1in}
\end{figure}

A LTW, shown in Fig. \ref{fig:spacetime}, consists of sequences of workflows which are indexed by a mobile user's \emph{location} and \emph{duration/time}.   It is represented more formally as follows:
\small
$$W(u_k)_{T}^{L} \triangleq (w(u_k)_{t_1}^{l_1},w(u_k)_{t_2}^{l_2},w(u_k)_{t_3}^{l_3},...,~w(u_k)_{t_n}^{l_n})$$
\normalsize
where $u_k$ is the $k^{th}$ mobile user and $w(u_k)_{t_n}^{l_n}$ is the user request workflow in location $l_n$ for time $t_n$.

So far we have modeled mobile users and their applications. In the next section we will model quality of service parameters for mobile applications.

\subsection{Quality of Services of Mobile Applications}

\begin{table}
\begin{center}
    \begin{tabular}{|l|p{14cm}|}
       \hline
       \textbf{Criteria} & \textbf{Definition}\\
       \hline
       $q_{price}(s_i,u^{l_i, t_j}_{k})$ & The price of using service $s_i$ when user $u_k$ is in location $l_i \in L$ and time $t_j$.\\
       \hline
       $q_{power}(s_i,u^{l_i, t_j}_{k})$ & The power consumed on user mobile device using $s_i$ when user $u_k$ is
       in location $l_i \in L$ and time $t_j$.\\
       \hline
       $q_{delay}(s_i,u^{l_i, t_j}_{k})$ & The delay of executing service $s_i$ when user $u_k$ is in location
       $l_i \in L$ and time $t_j$.\\
       \hline
     \end{tabular}
     \vspace{0.04in}
     \caption{QoS parameters that will be used in mobile cloud computing environment}
\label{tb:QoS}
\end{center}
\vspace{-.1in}
\end{table}

For mobile applications several Quality of Services (QoS) parameters such as \emph{delay}, \emph{power} and \emph{price} could be considered \cite{Qos_2009}. Table. \ref{tb:QoS} shows the quality of service parameters that we will be used in our mobile cloud computing environment. These QoS factors depend on user location and requested time. This is primarily due to the fact that communication link characteristics (Wi-Fi, 3G) vary based on user location and the time of the service. This in turn has an effect on the delay, power and price of the services and hence impacts the QoS. The delay of the service is considered as the difference between the time when a service is called (on the mobile device or cloud) and when the service is terminated. If the service on the cloud is being used we also account for the network delay (Wi-Fi or 3G). Power consumption of the service refers to the power consumed on mobile device to execute the service. If the service executes on the cloud, power consumed includes the power overheads of the network connection and data transfer related to that service. Finally, the {\it price} of the service is the actual price/cost to the end user of executing the service on the public cloud.

Table \ref{tb:AggrFormula} defines the QoS for the \emph{application workflow} based on the execution plan $\overrightarrow{x} \in \Gamma$. The QoS of a workflow is evaluated based on the QoS of its atomic services while taking into account the composition patterns \cite{Zeng_2004}. The QoS of a SEQ pattern is the sum of the QoSes of the constituent tasks for all QoS parameters (price, power, delay). In the case of the  AND pattern, that models parallel task flow, each of the QoS parameters is calculated independently. The price (power) of an AND workflow is the sum of the price (power) of the constituent tasks; the delay of the workflow is set to be the maximum delay of the parallel flows. In the XOR pattern, the maximum among the constituent values determines the QoS value all QoS types; for iterative tasks (i.e., structured as a LOOP), the QoS is determined by the number of executions of the service.

\begin{table}
\begin{center}
    \begin{tabular}{|>{\raggedright}p{1.4cm}|p{1.29cm}|p{1.29cm}|p{1.29cm}|p{1.27cm}|}
       \hline
       \textbf{QoS} & \textbf{SEQ} & \textbf{AND}& \textbf{XOR}& \textbf{LOOP}\\
       \hline
       $w(u_k)_{price}$ &  $\sum_{i=1}^n q_{price}^i$ &
       $\sum_{i=1}^n q_{price}^i$ & $\max_{i}~q_{price}$ & $q_{price}\times k$\\
       \hline
       $w(u_k)_{power}$ &  $\sum_{i=1}^n q_{power}^i$ &
       $\sum_{i=1}^n q_{power}^i$ & $\max_{i}~q_{power}$ & $q_{power}\times k$\\
       \hline
       $w(u_k)_{delay}$ &  $\sum_{i=1}^n q_{delay}^i$ &
       $\max_{i}~q_{delay}$ & $\max_{i}~q_{delay}$ & $q_{delay}\times k$\\
       \hline
     \end{tabular}
     \vspace{0.04in}
     \caption{Workflow QoS model}
\label{tb:AggrFormula}
\end{center}
\vspace{-.1in}
\end{table}

The extension of the workflow QoS to LTW Qos for single user $u_k$ could be done as:
\small
\begin{align}
\nonumber
&[W(u_k)_{T}^{L}]_{price} \triangleq \sum_{i=l_1, j=t_1}^{i=l_n, j=t_n}{ [w(u_k)_{j}^{i}]_{price}}\\
\nonumber
&[W(u_k)_{T}^{L}]_{power} \triangleq \sum_{i=l_1, j=t_1}^{i=l_n, j=t_n}{ [w(u_k)_{j}^{i}]_{power}}\\
\nonumber
&[W(u_k)_{T}^{L}]_{delay} \triangleq \sum_{i=l_1, j=t_1}^{i=l_n, j=t_n}{ [w(u_k)_{j}^{i}]_{delay}}\\
\nonumber
\end{align}
\normalsize

This concept could be easily expanded to the group of mobile users LTW QoS by summing up of each user experienced QoS. It could be formally defined as:
\small
\begin{align}
\nonumber
&[W(g_i)_{T}^{L}]_{price}  \triangleq \sum_{\forall u_k \in g_i}~\sum_{i=l_1, j=t_1}^{i=l_n, j=t_n}{ [w(u_k)_{j}^{i}]_{price}}\\
\nonumber
&[W(g_i)_{T}^{L}]_{power}  \triangleq \sum_{\forall u_k \in g_i}~\sum_{i=l_1, j=t_1}^{i=l_n, j=t_n}{ [w(u_k)_{j}^{i}]_{power}}\\
\nonumber
&[W(g_i)_{T}^{L}]_{delay}  \triangleq \sum_{\forall u_k \in g_i}~\sum_{i=l_1, j=t_1}^{i=l_n, j=t_n}{ [w(u_k)_{j}^{i}]_{delay}}\\
\nonumber
\end{align}
\normalsize

We require normalized values (for price, power, delay) that can be used to calculate the utility of the LTW of mobile users. \emph{This process is necessary while power, price and delay have different units like dollar, joule and second}. First, we will apply a normalization process \cite{Zeng_2004} for \emph{services}. We illustrate it in the context of price, but is easily generalized to power and delay.

\begin{itemize}
  \item \small $Price^{max}(\chi_{Fi})$ \normalsize : The maximum price of the services that could realize function $Fi$.
  \item \small $Price^{min}(\chi_{Fi})$ \normalsize : The minimum price of the services that could realize function $Fi$.
  \item For each services \small $s \in \chi_{Fi}$ \normalsize the normalized price could be defined as:
  \end{itemize}
\small
\begin{align}
\nonumber
\|s_{price}\| \triangleq
\left\{
\begin{array}{l l}
    \frac{Price^{max}(\chi_{Fi})- s_{price} }{ Price^{max}(\chi_{Fi})-Price^{min}(\chi_{Fi})}&\hspace{-0.05in}
    \quad\text{$Price^{max}(\chi_{Fi})$}\\
    &\hspace{-0.1in} \quad\text{$\neq Price^{min}(\chi_{Fi})$}\\
    1 & \quad \text{else}\\
\end{array} \right.
\nonumber
\end{align}
\normalsize

For each services $s \in \chi_{Fi}$ the total normalized QoS is defined as: \small $ \|s\| \triangleq [{\|s_{pow}^2\|+\| s_{price}^2\|+\|s_{delay}^2\|}]^\frac{1}{2}$ \normalsize. In general the \emph{higher} the $\|s\|$ is, the better the QoS/performance (small delay, power consumption and price) of the service.

The next step in normalization process is to extend it to the \emph{workflow} $w$. Again, we illustrate this step using the price (trivially extended to power and delay).
\begin{itemize}
  \item \small $ C_{price}^{max} $ \normalsize : The total price of the services in workflow when the most expensive services are selected.
      \vspace{0.02in}
  \item \small $ C_{price}^{min} $ \normalsize : The total price of the services in workflow when the cheapest services are selected.
      \vspace{0.02in}
  \item \small $ \|w(u_{k})_{price}\|$ \normalsize : \emph{Normalized price} of the workflow with specific service plan \small $\overrightarrow{x} \in \Gamma$ \normalsize is defined as:
\vspace{-0.04in}
\begin{align}
\nonumber
\small
\|w(u_{k})_{price}\| \triangleq
\nonumber
\left\{
  \begin{array}{l l}
  \small
    \frac{C_{price}^{max}-w_{price}(u_{k})}
    {C_{price}^{max}-C_{price}^{min}} & \quad \small \hspace{-0.15in} \text{$C_{price}^{max} $}\\
    & \hspace{-0.15in}\neq \hspace{-0.15in} \quad \text{$C_{price}^{min}$}\\
    1 & \quad \text{else}\\
  \end{array} \right.
\end{align}
\end{itemize}

The same procedure could be done for the LTW and Group-Based LTW. As an example we show for LTW but could be easily extended to Group-based LTW:

\begin{itemize}
  \item \small $ [C_T^{L}]_{price}^{max} $ \normalsize : The total price of the services in LTW when the most expensive services are selected.
      \vspace{0.02in}
  \item \small $ [C_T^{L}]_{price}^{min} $ \normalsize : The total price of the services in LTW when the cheapest services are selected.
      \vspace{0.02in}
  \item \small $ \|[W(u_k)_{T}^{L}]_{price}\|$ \normalsize : \emph{Normalized price} of the space-time workflow with specific service plan $\overrightarrow{x} \in \Gamma$ is defined as:
\vspace{-0.1in}
\small

\begin{align}
\nonumber
\|[W(u_k)_{T}^{L}]_{price}\| \triangleq~~~~~~~~~~~~~~~~~~~~~~~~~~~~~~~~~~~~~~~~~~~~~~~~~~~~~~\\
  \nonumber
  \vspace{0.05in}
  \nonumber \left\{
  \begin{array}{l l}
  \small
    \frac{[C_T^{L}]_{price}^{max}-[W(u_k)_{T}^{L}]_{price}}
    {[C_T^{L}]_{price}^{max}-[C_T^{L}]_{price}^{min}} & \small \hspace{0.1in} \text{$[C_T^{L}]_{price}^{max}$}~~~~~~~~~~~~~~\\ & \neq \text{$[C_T^{L}]_{price}^{min}$}~~~~\\
    1 & \quad \text{else}\\
  \end{array} \right.
\end{align}
\end{itemize}
\normalsize

LTW and QoS give us a formal and solid framework which we could study the performance of the mobile applications on the cloud computing environment. The next important concept that we should consider is the \emph{Utility Function} which models formally the general performance of the system.

Different utility functions could be defined that consider the service providers benefits, mobile users benefits or both, but in this paper our main concern is benefit of mobile users or group of mobile users. We define the \textbf{mobile users utility} as:

\vspace{-0.2in}
\small
\begin{align}
\nonumber
\digamma_{mobile} \triangleq ~~~~~~~~~~~~~~~~~~~~~~~~~~~~~~~~~~~~\\
\nonumber
\frac{1}{n}\sum_{u_k}\min \{ \|[W(u_k)_{T}^{L}]_{price}\|, \|[W(u_k)_{T}^{L}]_{power}\|,\\
\nonumber
\small
\|[W(u_k)_{T}^{L}]_{delay}\|\}~~~~~~~~~~~~~~~~~~~~~~~\\
\nonumber
\end{align}
\normalsize
\vspace{-0.3in}

Intuitively this function results the average of minimum saving of price, power and delay of mobile users as the mobile users benefits.

We extend this single user utility function to \textbf{mobile group $g_i$ utility} as:

\vspace{-0.2in}
\small
\begin{align}
\nonumber
\digamma_{mobile}^{g_i} \triangleq ~~~~~~~~~~~~~~~~~~~~~~~~~~~~~~~~~~~~\\
\nonumber
\frac{1}{|g_i|}\sum_{u_k \in g_i}\min \{ \|[W(u_k)_{T}^{L}]_{price}\|, \|[W(u_k)_{T}^{L}]_{power}\|,\\
\nonumber
\small
\|[W(u_k)_{T}^{L}]_{delay}\|\}~~~~~~~~~~~~~~~~~~~~~~~\\
\nonumber
\end{align}
\normalsize
\vspace{-0.35in}

This function results the average of minimum saving of price, power and delay of mobile group users as the group benefits.

By combining the utility function and \emph{system constraints} we end up with the following two optimization problems for the service allocation on mobile cloud computing. The first one is for single users optimal service allocation and states as:

\small
\begin{align}
\nonumber
&\hspace{-0.2in} \max \digamma_{mobile}\\
\nonumber
& st:\\
\nonumber
&\frac{1}{n}\sum_{u_k} [W(u_k)_{T}^{L}]_{price} \leq B_{price}\\
&\frac{1}{n}\sum_{u_k} [W(u_k)_{T}^{L}]_{power} \leq B_{power}\\
\nonumber
&\frac{1}{n}\sum_{u_k} [W(u_k)_{T}^{L}]_{delay} \leq B_{delay}\\
\nonumber
&\kappa \leq Cap(Local~Cloud),~\kappa \leq n\\
\nonumber
&\kappa \triangleq \text{Number of Mobile users using services on local cloud.}\\
\nonumber
&\forall u_k \in \{u_1, u_2,..., u_n\}\\
\nonumber
\end{align}
\normalsize

The first, second and third constraints say that the user spent price, consumed power and experienced delay should be less than a limit. The final constraints are the local cloud constraint which could only accept a limited number of mobile users requests.

The second optimization problem is related to group of mobile users. In this problem our goal is to optimize the average QoS of group members. It is stated as:

\small
\begin{align}
\nonumber
&\hspace{-0.2in} \max \frac{1}{|G|}\sum_{g_i \in G}\digamma_{mobile}^{g_i}\\
\nonumber
& st:\\
\nonumber
&\frac{1}{|g_i|}\sum_{g_i \in G} [W(g_i)_{T}^{L}]_{price} \leq B_{price}^{g_i}\\
&\frac{1}{|g_i|}\sum_{g_i \in G} [W(g_i)_{T}^{L}]_{power} \leq B_{power}^{g_i}\\
\nonumber
&\frac{1}{|g_i|}\sum_{g_i \in G} [W(g_i)_{T}^{L}]_{delay} \leq B_{delay}^{g_i}\\
\nonumber
&\kappa \leq Cap(Local~Cloud),~\kappa \leq n\\
\nonumber
&\kappa \triangleq \text{Number of Mobile users using services on local cloud.}\\
\nonumber
&\forall u_k \in \{u_1, u_2,..., u_n\}\\
\nonumber
&\forall g_i \in \{g_1, g_2,..., g_{|G|}\}\\
\nonumber
\end{align}
\normalsize

As before the first, second and third constraints say that the average group spent price, consumed power and experienced delay should be less than a limit.

Both of the mentioned problems are NP-Hard while Knapsack is the special case of it . In the next section we will propose a heuristic to solve this problem.

\section{ Simulated Annealing Based Heuristic for Resource Allocation in the Tiered Cloud}\label{ProposedAlgorithm}
We extend our previous works \cite{MAPCloud_2012}, \cite{Music_2013} and develop \textbf{MuSIC} (\textbf{M}obility-Aware \textbf{S}ervice Allocat\textbf{I}on on \textbf{C}loud), an efficient heuristic for tiered-cloud service allocation which takes into consideration user mobility information and supports both type of mobile applications mentioned in the previous section. MuSIC algorithm is a greedy heuristic that generates a near-optimal solution to the tiered cloud resource allocation problem using a simulated annealing-based approach, which has been shown to be an efficient heuristic for knapsack problem \cite{Knapsack_1998}. A simulated annealing approach typically starts out with an initial solution in the potential solution space and iteratively refines this to generate increasingly improved solutions. It uses a randomized approach to increase the diversity of service selection \cite{simulated_anealing}.

Table \ref{tb:MUSIC} contains pseudo code for the MuSIC algorithm. While MuSIC uses simulated annealing as the core approach in selecting and refining service selection; custom policies have been designed to make it efficient for the tiered cloud architecture with mobile applications. Given a set of users or group of users with their corresponding LTWs $W(*)_{T}^{L}$, a constraints set $\overrightarrow{C}$, $SingleUserFlag$ which indicates that the current run of the MuSIC is for single user or group of users, $S$ which is the service DB and
$max_{iter}$ which shows the maximum iteration of simulated annealing. Based on $SingleUserFlag$  MuSIC starts by computing the center of mobility $l^{u_k}_{cm}$ of each user $u_k$ or $l^{g_i}_{cm}$ of each group $g_i$. Intuitively it is a location in the single/group mobile user's trajectory where much of the time is spent; the general goal is to select services near that location. MuSIC then uses the service selection function $Find_{Service}( W(*)_{T}^{L}, \overrightarrow{C}, l^{*}_{cm})$  that returns the list of services near the user center of mobility $l^{u_k}_{cm}$ or group user center of mobility $l^{g_i}_{cm}$, which can realize the LTW and satisfy the required constraints.

In lines 4 and 18 the utility functions $\digamma_{mobile}$ or $\digamma_{mobile}^{g_i}$ of this solution are computed. Following this, the MuSIC algorithm will enter a loop which is the main core for the simulated annealing based algorithm. The difference between the initial value of the  $\digamma_{mobile}$ or $\digamma_{mobile}^{g_i}$ function and current computed value of  $\digamma_{mobile}$ or $\digamma_{mobile}^{g_i}$ function is extracted in lines 8 and 22. If it is positive, it will be then considered as the new service list; if negative, it may still be retained with some probability and the algorithm will enter the next iteration. The while loop is eventually terminated when the number of iterations exceeds a limit $max_{iter}$. After the iterations are done the final utility and service set will be returned as the solution.

\begin{table}[t]
\begin{center}
\begin{tabular}{|p{8.25cm}|}
\hline
$MuSIC~( W(*)_{T}^{L},~SingleUserFlag,~S,~\overrightarrow{C},~max_{iter})$\\
~~~~~~~~~~~~~~~~~\\
$W(*)_{T}^{L}$:~$u_k$ or $g_i$ LTW; $SingleUserFlag$: true if LTW is for single user false if it is for group of users~; $S$: Service Set DB; $\overrightarrow{C}$: Constraint Vector; $max_{iter}$: Simulated Annealing Number of Iteration.
~~~~~~~~~~~~~~~~~\\
\\
Begin\\
\\
(1)~\textbf{if} $SingleUserFlag=True$.\\
(2)~~~~Compute $l^{u_k}_{cm}$.\\
(3)~~~~$Candidate_{Service}=Find_{Service}( W(u_k)_{T}^{L}, \overrightarrow{C}, l^{u_k}_{cm})$\\
(4)~~~~$Util_0=Compute_{\digamma_{mobile}}(Candidate_{Services})$\\
(5)~~~~For j=1 to $max_{iter}$ do\\
(6)~~~~~~~$Candidate_{Services}=Find_{Service}(W(u_k)_{T}^{L}, \overrightarrow{C}, l^{u_k}_{cm})$\\
(7)~~~~~~~$Util_1= Compute_{\digamma_{mobile}}(Candidate_{Services})$\\
(8)~~~~~~~$\Delta$ =$Util_1-Util_0$\\
(9)~~~~~~~~~~If $\Delta >$ 0\\
(10)~~~~~~~~~~~~~$Util_0=Util_1$\\
(11)~~~~~~~~~Else\\
(12)~~~~~~~~~~~~Replace $Util_0=Util_1$ when $exp(max_{iter})~\geq~U[0,1]$\\
~~~~~~~~~~~~~~/* U[0,1] means the uniform distribution function */\\
(13)~~~~~~~~~End if\\
(14)~~~End for\\
(15)~\textbf{else}\\
(16)~~~~Compute $l^{g_i}_{cm}$.\\
(17)~~~~$Candidate_{Service}=Find_{Service}( W(g_i)_{T}^{L}, \overrightarrow{C}, l^{g_i}_{cm})$\\
(18)~~~~$Util_0=Compute_{\digamma_{mobile}^{g_i}}(Candidate_{Services})$\\
(19)~~~~For j=1 to $max_{iter}$ do\\
(20)~~~~~~~$Candidate_{Services}=Find_{Service}(W(g_i)_{T}^{L}, \overrightarrow{C}, l^{g_i}_{cm})$\\
(21)~~~~~~~$Util_1= Compute_{\digamma_{mobile}^{g_i}}(Candidate_{Services})$\\
(22)~~~~~~~$\Delta$ =$Util_1-Util_0$\\
(23)~~~~~~~~~~If $\Delta >$ 0\\
(24)~~~~~~~~~~~~~$Util_0=Util_1$\\
(25)~~~~~~~~~Else\\
(26)~~~~~~~~~~~~Replace $Util_0=Util_1$ when $exp(max_{iter})~\geq~U[0,1]$\\
~~~~~~~~~~~~~~/* U[0,1] means the uniform distribution function */\\
(27)~~~~~~~~~End if\\
(28)~~~End for\\
\\
(29)~~~Return $Candidate_{Service}, Util_0$\\
\\
End\\
\\
\hline
\end{tabular}
\end{center}
\caption{ MuSIC Algorithm Pseudo code }
\label{tb:MUSIC}
\vspace{-.3in}
\end{table}

The main core of MuSIC is the $Find_{Service}$ function which returns the candidate service set for $W(u_k)_{T}^{L}$ or $W(g_i)_{T}^{L}$. There are two intuitions behind this function. First of all it is known that services in close proximity to the user usually provide better QoS performance in terms of delay and power consumption. Secondly using services with high total QoS will increase system utility.
In $Find_{Service}$ module, we facilitate a better initial solution by veering the service selection towards those services in close proximity to the user. This is realized by storing the services in broker directory service/registry using a structure that enable efficient retrieval of  nearby user services. Specifically, we store services using an \emph{R-tree} based data structure \cite{DB_2010}. Such an R-tree based data structure has been used for storing geometrical data and has been shown to enable efficient search, insertion, deletion and updates.

\begin{figure}[b]
\begin{center}
\includegraphics[width=4in,height=3in]{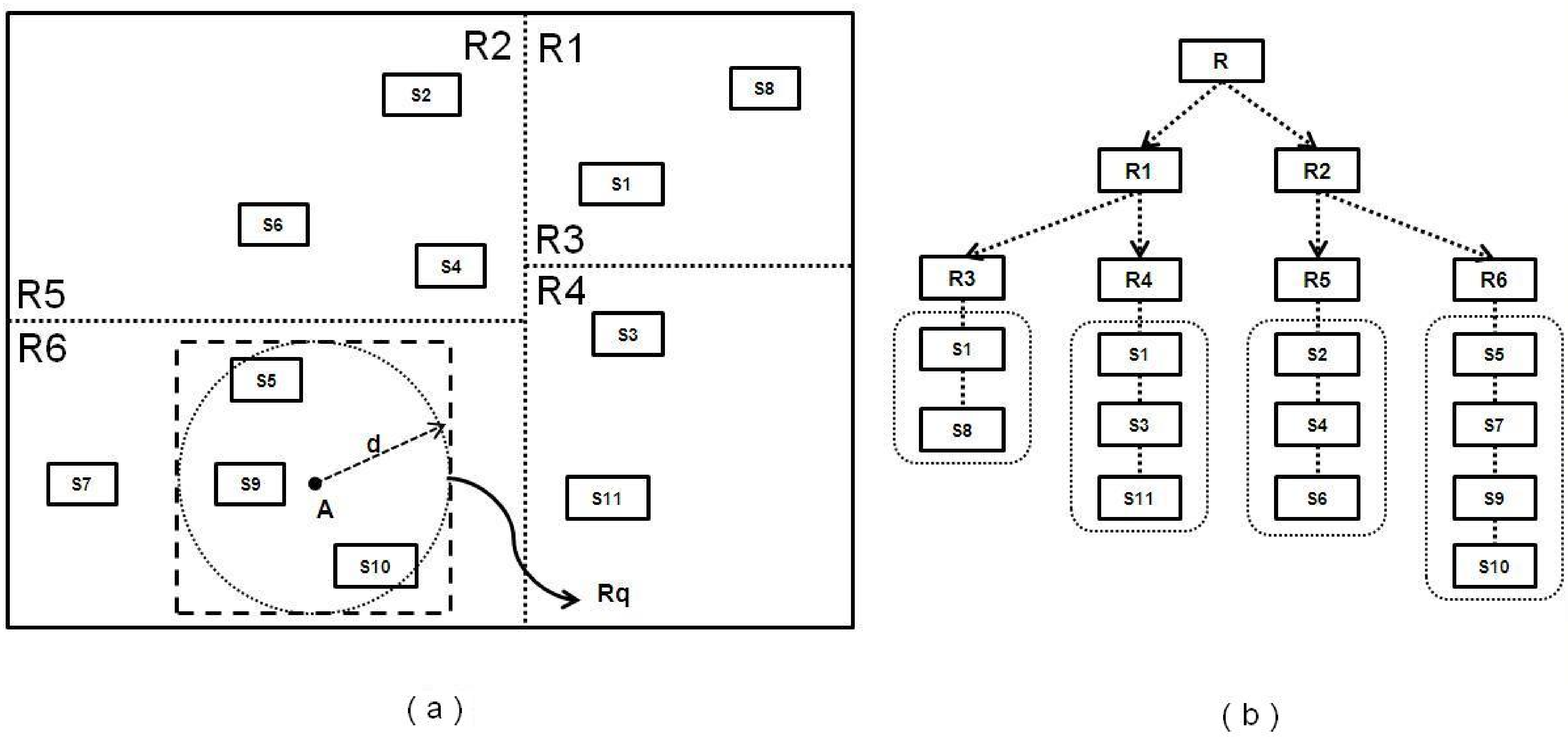}
\caption{R-Tree Data Structure: ( a ) Partitioning the 2-D space into rectangles ( b ) R-Tree structure
of services}
\label{fig:rtree}
\end{center}
\vspace{-.1in}
\end{figure}

Fig. \ref{fig:rtree} shows a sample R-Tree data structure for services. The R-tree structure splits the search space into hierarchically nested, and possibly overlapping, minimum bounding rectangles. We next illustrate how efficient retrieval of services near a user can be realized using an R-Tree data structure. As an example suppose we are interested in query "\emph{Retrieve all services in distance $d$ of point A} " as shown in Fig. \ref{fig:rtree} (a). The system will create a minimum rectangle that contains a circle with center $A$ and radius $d$. This rectangle is called $R_q$ in Fig. \ref{fig:rtree} (a). Then it will search and find all overlapping rectangle with $R_q$ which is in our case is $R_6$ and retrieve all services in $R_6$. In best case if the number of records in data base is $n$ then with using R-Tree structure we could retrieve our records in $O(log(n))$.

Table \ref{tb:MUSIC_find_single} illustrates the $Find_{Service}$ routine. It starts with a candidate set of services, $Candidate_{Services}$ within a threshold distance $d=d_{th}$ from the $l^{*}_{cm}$. If they satisfy the constraints then it starts a loop in line 6. Loop starts by sorting services based on \textbf{total QoS} from small to large $(\|s_1\|,\|s_2\|,...,\|s_n\|)$. It then makes normalized service based on that vector, $\overrightarrow{v}=(\frac{\|s_1\|}{\sum_{i}\|s_i\|},...,\frac{\|s_n\|}{\sum_{i}\|s_i\|} )$. our goal is to select one of the services based on its normalized value which could \emph{balance} the service selection for all users or groups. We then generate a random number between in $[0,1]$.
If $a \in [\frac{1}{sum}\sum_{i=0}^{i=j}\|s_{i}\|,\frac{1}{sum}\sum_{i=0}^{i=j+1}\|s_{i}\|]$ then we select $s_{j+1}$ for $W(*)_{T}^{L}$.
As an example suppose that we are looking for \emph{MPEG-TO-FLASHVIDEO} service. $Find_{Service}$ finds 3 different services $s_{1},s_{2},s_{3}$. We then sort this list based on \textbf{total QoS} from small to large $(S_{1},S_{2},S_{3})=(s_{2},s_{1},s_{3})$. Then we make a normalized vector $(\|S_{1}\|=0.2,\|S_{2}\|=0.3,\|S_{3}\|=0.5)$. Our goal is to select one of these services. If we choose greedy approach we then select $S_{3}$, which have the \emph{higher total QoS}. But we use \emph{randomized strategy} to ad more diversity in service selection. If generate random number $a$ in $[0,1]$ and suppose it is 0.35. Then we will select $S_{2}$ while $a$ is in $[S_{1}=0.2,S_{1}+S_{2}=0.5]$.

In the next section we will present system prototyping and profiling results.

\begin{table}
\begin{center}
\begin{tabular}{|p{8.25cm}|}
\hline
$Find_{Service}(W(*)_{T}^{L}, \overrightarrow{C}, l^{*}_{cm})$\\
~~~~~~~~~~~~\\
We assume that the directory service database contains information
on the normalized QoS of the service with \emph{\textbf{R-Tree indexing}}.
~~~~~~~~~~~~\\
\\
$W(*)_{T}^{L}$: $u_k$ or $g_i$ Location-Time Work Flow, $S$: Service Set DB, $l^{*}_{cm}$: $u_k$ or $g_i$ center of mobility,
const $d_{th}:$ Threshold Distance, const $d_r:$ The increase amount of distance,
const $it:$ Maximum number of iteration
~~~~~~\\
\\
$Begin$\\
\\
(1)~$i$ = 0;\\
(2)~$\textbf{while}$ ($i$ $<$ $it$)\\
~~~~$\textbf{begin}$\\
(3)~~~$d=d_{th}+i*d_r$\\
(4)~~~$Candidate_{Services}$=Retrive the related services according to $W(*)_{T}^{L}$ \\~~~~~~~in distance $d$ of $l^{*}_{cm}$.\\
(5)~~~~~$\textbf{if}$ $Candidate_{Services}$ contains all of the needed services and satisfies\\~~~~~~~the constraints:\\
(6)~~~~~~~~~~~~$\textbf{foreach}$~$s_k \in Candidate_{Services}$ do:\\
~~~~~~~~~~~~~~~~~~make a sorted list according to \textbf{normalized total QoS} \\~~~~~~~~~~~~~~~~~~~from small to large $(\|s_1\|,\|s_2\|,...,\|s_n\|)$.\\
(7)~~~~~~~~~~~~~~~~Make a vector $\overrightarrow{v}$ using total normalized QoS according \\~~~~~~~~~~~~~~~~~~~to: $sum=\sum_{i}\|s_i\|, \overrightarrow{v}=(\frac{\|s_1\|}{sum},...,\frac{\|s_n\|}{sum} )$.\\
(8)~~~~~~~~~~~~~~~~generate random number $a \in [0,1]$.\\
(9)~~~~~~~~~~~~~~~~$if$ $a \in [\frac{1}{sum}\sum_{i=0}^{i=j}\|s_{i}\|,\frac{1}{sum}\sum_{i=0}^{i=j+1}\|s_{i}\|]$ then
\\~~~~~~~~~~~~~~~~~~~select $s_{j+1}$ for $W(*)_{T}^{L}$.\\
(10)~~~~~~~~~~~$\textbf{endFor}$\\
(11)~~~~~~~~~~return the service set\\
(12)~~~~~$\textbf{else}$\\
(13)~~~~~~~~$i=i+1$\\
(14)~~~~~~~~increase the search radios to $d=d_{th}+i*d_r$\\
(15)~$\textbf{end~while}$\\
\\
$End$\\
\\
\hline
\end{tabular}
\end{center}
\caption{ MuSIC $Find_{Service}$ Algorithm Pseudo code }
\label{tb:MUSIC_find_single}
\vspace{-.3in}
\end{table}

\section{System Prototyping and Profiling} \label{Evaluation}

\begin{figure}[b]
\begin{center}
\includegraphics[width=4in,height=2.25in]{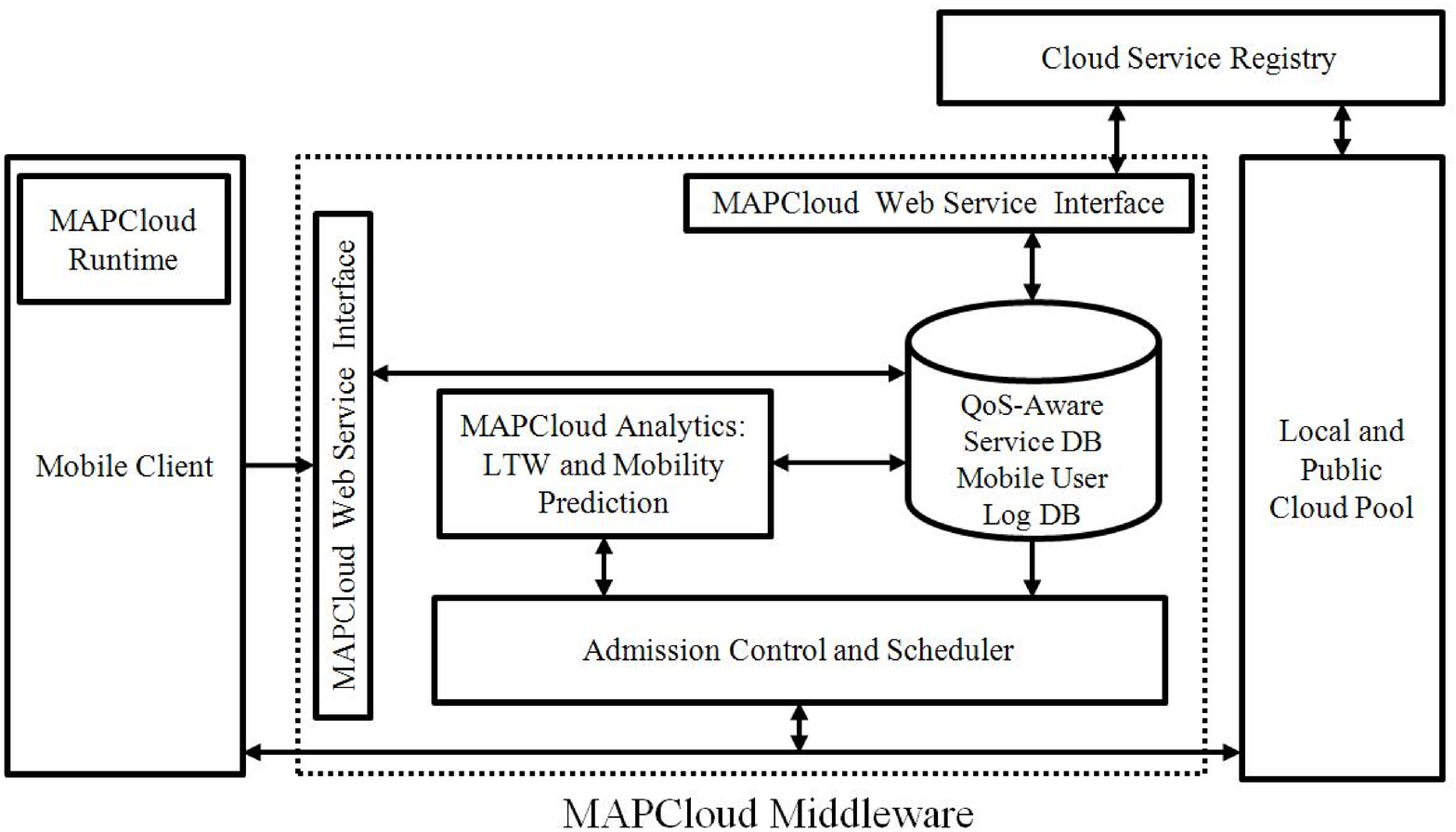}
\caption{ Middleware Service Architecture }
\label{fig:midware}
\end{center}
\vspace{-.1in}
\end{figure}

We extend MAPCloud middleware to support LTW and MuSIC \cite{MAPCloud_2012}, \cite{Music_2013}. Fig. \ref{fig:midware} illustrates the general architecture of MAPCloud platform with the key modules describe below:

\textsl{Mobile User Log DB and QoS-Aware Service DB}: The first one contains unprocessed user data log such as mobile service usage, location of the user, user delay experience of getting the service, energy consumed on user mobile device, etc. The second one contains the service lists on local and public cloud and their QoSes in different locations.

\textsl{MAPCloud Analytic}: This module processes mobile user Log DB and updates QoS-aware cloud service DB based on user experience and LTW.

\textsl{Admission Control and Scheduling}: This module is responsible for optimally allocate services to admitted mobile users based on MuSIC.

The operational flow through this module is simple -a user requested mobile application is forwarded to the MAPCloud. If admitted (based on service availability), the scheduler module will compute and determine the best allocation of services using the MuSIC algorithm. The scheduler modules consult the QoS-Aware Cloud DB and MAPCloud Analytic. The service plan is returned back contains URL of each services in application LTW.

To study MuSIC performance OCR+Speech (OCRS), video streaming and transcoding (VS) and multimedia file sharing (MFS) applications have been developed as the rich mobile applications. In the first application the user takes a picture of the text page and the application will return a file which contains the spoken text. The second application is video streaming and transcoding application in which the video clip is streamed to the mobile users. The third application (MFS) is group based application. In this application mobile users share multimedia data. They could edit data, watch videos and upload/download multimedia files.

For the mentioned mobile applications different services has been extracted such as image filtering, noise cancelation, transcoding, etc. We measure the delay and power consumption of services in different situation for both local and public cloud. The following procedure has been used for measuring power, delay and price on local and public cloud for different services:

\vspace{0.081in}
\textsl{Delay Profiling:}
\vspace{0.081in}

\begin{figure}
\begin{center}
\includegraphics[width=4.5in,height=4.2in]{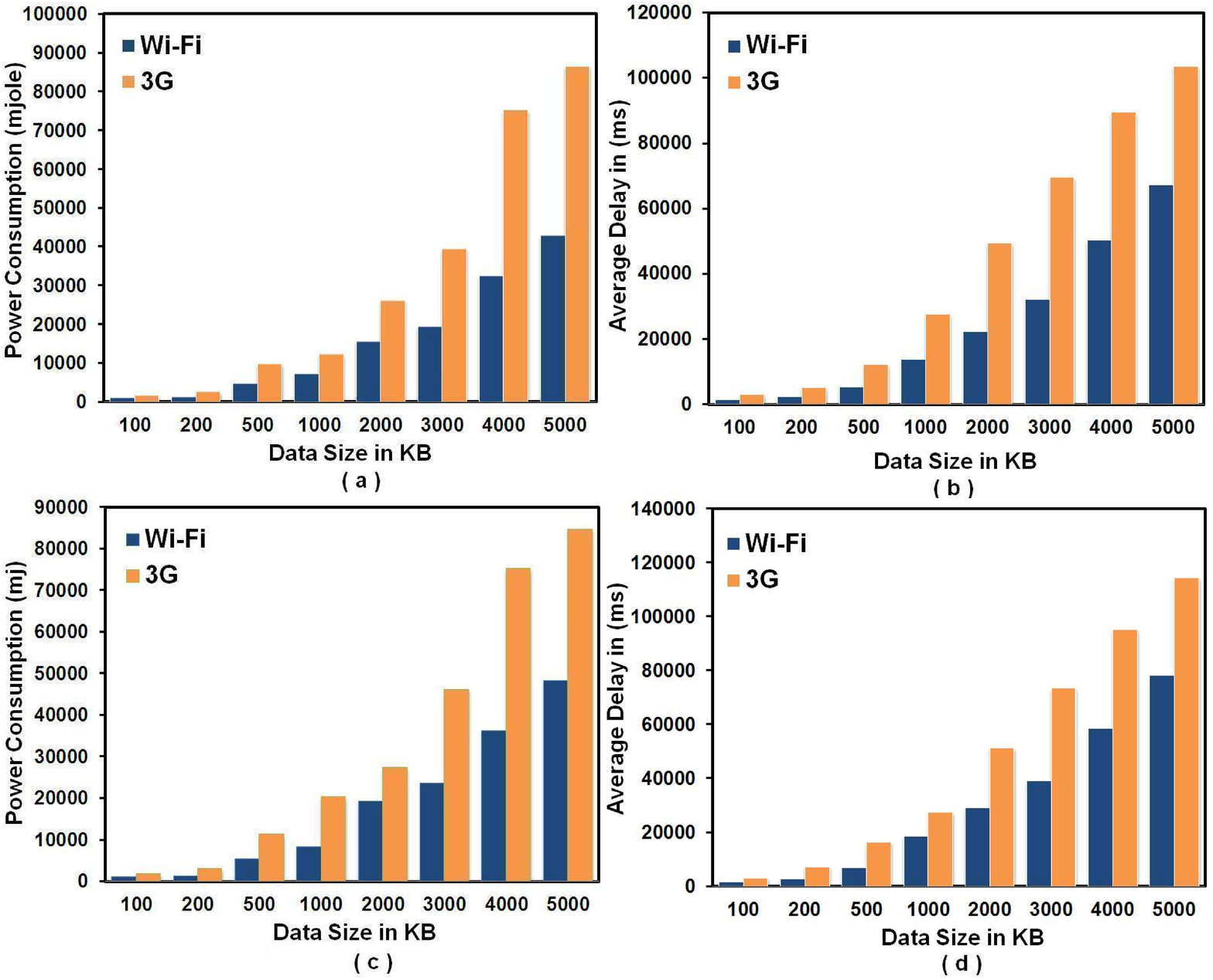}
\caption{ Averaged Delay (in ms) and power consumption (in mjole) of different wireless network types regarding to data size when using local cloud ( Fig. a and b) and Amazon Public Cloud (Fig. c and d).}
\label{fig:powerdelaypublic}
\end{center}
\vspace{-.1in}
\end{figure}

Four different delays have been considered as:
\begin{itemize}
\item $D_p^{s_i}$: The delay caused by processing on cloud. We define an average processing per \emph{100KB} of data for each $s_i$. The averaging has been done on large and different number of services on local and public clouds to get the $D_p^{s_i}$.

\item $D_{wifi}$: The delay of using Wi-Fi as the communication link to transfer data to cloud (or download from cloud ). The $D_{wifi}$ is defined as the average delay of transmitting 100KB of data over Wi-Fi. Different packet sizes have been considered to transfer data from mobile device to local or public clouds (from 100KB up to 5MB).
Fig.\ref{fig:powerdelaypublic} (b) shows the average delay of transmitting/recieving data from Android G2 to local cloud using Wi-Fi and 3G with different data size. For example for typical 2Mb of file size the average Wi-Fi delay is about 220 ms. This delay is longer when using public cloud as shown in figure Fig. \ref{fig:powerdelaypublic} (d). In this case for 2Mb of file size the average Wi-Fi delay is about 240 ms.

\item $D_{3g}$: The delay of using 3G as the communication link to transfer data to cloud (or download from cloud ). We define $D_{3g}$ as the average delay of transmitting 100KB of data over $D_{3g}$. We have used different packet sizes to transfer from mobile device to cloud and then averaged among all of them (from 100KB up to 5MB).
As it is shown in Fig.\ref{fig:powerdelaypublic} (b)  for typical 2Mb of file size transfer using local cloud the delay is 4426ms using 3G. It becomes 5128ms for public cloud.

\item $D_{ic}$: The intercloud delay is considered as the delay of transmitting or receiving data among clouds.
\end{itemize}

The above delays are considered as the main delay source in the system. If two services in \emph{sequential pattern} in workflow should be implemented in different locations, the communication delay would be considered as the service delay. In summary this could be modeled as:

\vspace{-.023in}
\small
$$ Service~Delay=D_p^{s_i}+D(D_{wifi}, D_{3g}, D_{ic})$$
\normalsize
In this formula function $D$ represents the aggregated delay caused by communication link.

\vspace{0.081in}
\textsl{Power Profiling:}
\vspace{0.081in}

While the power consumption on user device is important, we have considered the following parameters as:

\begin{itemize}
\item $Pow_{dev}$: Consumed power of services on device. The PowerTutor \cite{power_tutor_2010} has been used to measure the power consumption of some services on Android G2.

\item $Pow_{wifi}$ consumed power of the device when transmitting 100KB of data using Wi-Fi. We define $Pow_{wifi}$ as the average power consumption of 100KB of data over Wi-Fi. We have used different packet sizes to transfer from mobile device to cloud and then averaging among all of them (from 100KB up to 5MB). Fig. \ref{fig:powerdelaypublic} (a) and (c) shows the average power consumption of transmitting/recieving data from Android G2 to local and public cloud with different data size. For example for typical 2Mb of file size the average Wi-Fi power consumption is 15435 mjole. This
power consumption is more when using public cloud as shown in figure Fig. \ref{fig:powerdelaypublic} (c). For 2Mb of file size the average Wi-Fi power consumption is 19345 mjole.

\item $Pow_{3G}$ consumed power of the device when transmitting 100KB of data using 3G. We defined $Pow_{3G}$ as the average power consumption of 100KB of data over 3G. We have used different packet size (different file size) to transfer from mobile device to cloud and then averaging among all of them (from 100KB up to 5MB). As it is shown in Fig. \ref{fig:powerdelaypublic} (a) and (c) for typical 2Mb of file size the average 3G power consumption is 26156 mjole for local cloud . It becomes 27345 mjole when using public cloud.
\end{itemize}

The above power consumption sources are considered as the main source in the system.
Again as mentioned above, if two services in \emph{sequential pattern} in workflow should be implemented in different locations, we would consider the power consumption of communication link as their service power. In summary this could be modeled as:

\vspace{-.023in}
\small
$$ Service~Power~Consumption=Pow_{dev}+P(Pow_{wifi}, Pow_{3G})$$
\normalsize
In this formula function $P$ represent the aggregated power consumption caused by different communication links on mobile device.

\vspace{0.081in}
\textsl{Price profiling:}
\vspace{0.081in}

Amazon pricing model has been used for the services on cloud and T-mobile data service plan as the price of using 3G for sending and receiving data. For Amazon EC2 \cite{Amazon_WS} the \emph{large instance} has been used in simulation (\$0.52 per hour). For measuring the price of each services on Amazon EC2, we assign different tasks with different data size. We then average over all data to have the price of each services for 100KB of data. We have use Amazon S3 services for data storage.
It has \$0.140 per GB storage \$0.1 per GB data transfer. \emph{Wowza} \emph{media streaming server} \cite{Wowza} with 0.15\$ per hour has been used as the video streaming on public cloud. To measure the 3G price the T-Mobile \cite{TMobile} data plan has been used (40\$ per 2GB/month). We considered that local cloud services and Wi-Fi connection are free.

In the next section we will present the simulation and performance results.

\section{Simulation Results} \label{SEvaluation}

Simulation platform is used to test the performance and scalability of the proposed system architecture and algorithms. In particular, we used MATLAB and CloudSim \cite{cloudsim_2011}, an open source cloud simulator which supports modeling of data centers, virtual machines and resource provisioning policies in a cloud computing environment. The experimental result obtained by profiling real applications in the prototype has been used to tune the simulation environment.

The basic simulation setup models a region with 225 cells (15 $\times$ 15). Local clouds have valid Wi-Fi in 6 cells around and there exists 3G connectivity in whole region. A LAN provides a backbone for local cloud connectivity and data transfer. We used two important mobility model in our simulation environment one is \textbf{Random Waypoint} (RW) and \textbf{Manhattan} models \cite{Mobility_2002}. Manhattan mobility model is mainly used for the movement in urban area, where the streets are in an organized manner.

We used the 15 $\times$ 15 grid size in our simulation. In our simulation we used the speed range in [1m/s, 10m/s]. we combine these two models in our simulation environment in a sense that 50\% of mobile users have RW model the remaining have manhattan model. In our simulation environment we assumed that half of the time mobile users are using \emph{OCRS} and half of time they are using VS applications for testing single users application. To test the performance of MuSIC for group based and collaborative applications (\emph{for simplicity we call it} \emph{G-MuSIC} during rest of simulation section ) we considered \emph{MFS}. In this scenario we consider the different groups size with different mobility models.

We set the maximum number of MuSIC iterations to 20. In our experiments, we varied data sizes which were uniformly distributed from [1Mb, 5Mb]. Each simulation results is the average of 15 runs. We test the performance of the system based on different number of users, different number of public and local cloud instances and uncertainty in prediction of mobile users' LTW. For example 10\% uncertainty in LTW consists of 10 sub-workflow means that in average one from 10 is not predicted correctly (error in prediction of user's location or requested service). In our simulation we considered the uncertainty in the range of [0\%-30\%].

\vspace{0.07in}
\textsl{\textbf{MuSIC Optimality Study :}}
\vspace{0.07in}

\begin{figure}[t]
\begin{center}
\includegraphics[width=4.35in,height=4.35in]{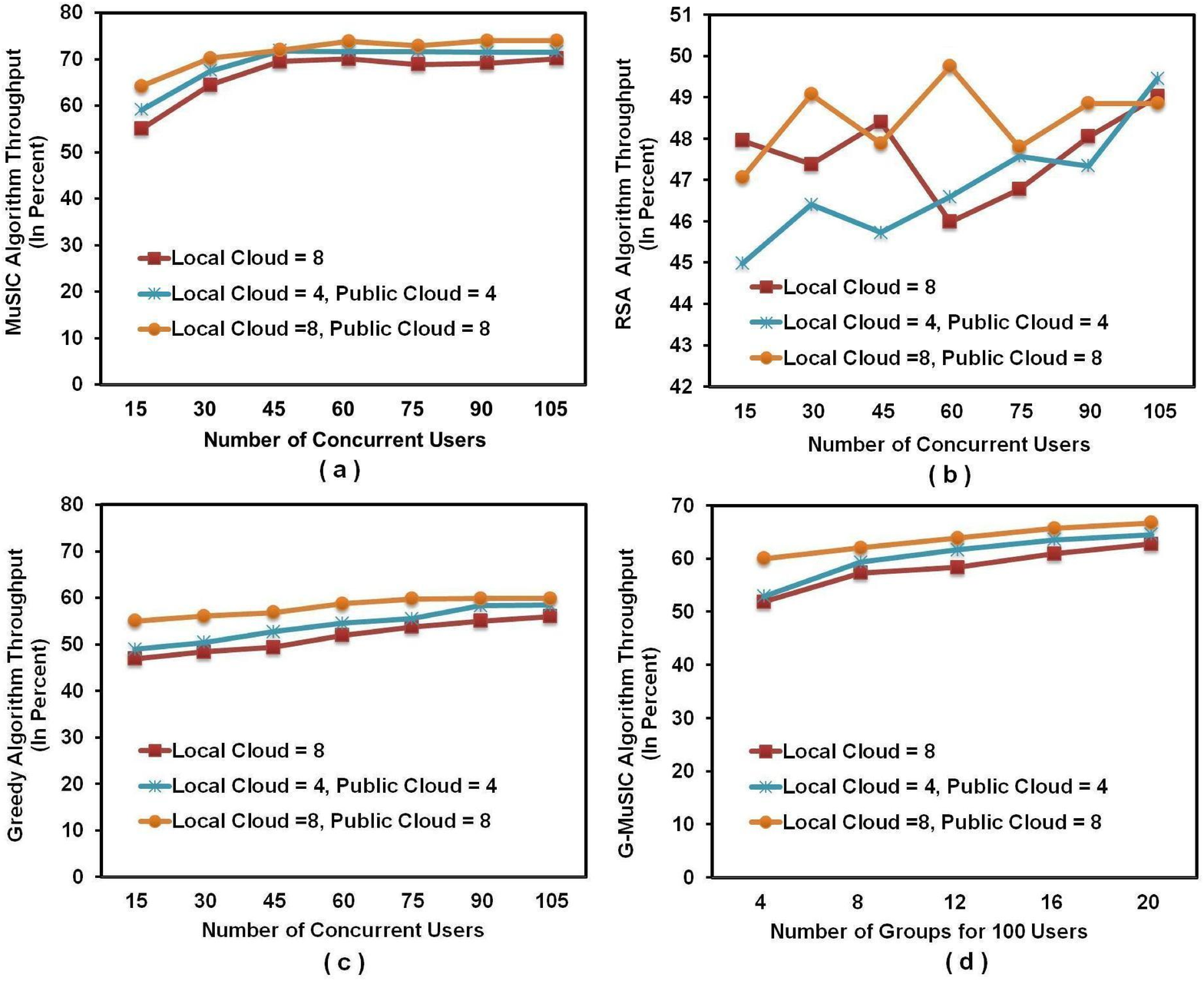}
\caption{ MuSIC, RSA, Greedy and G-MuSIC algorithms average throughput with uncertainty in the range of [0\%,30\%]}
\label{fig:MUSICRSA_Optimality}
\end{center}
\vspace{-.1in}
\end{figure}

To measure MuSIC and G-MuSIC optimality we compare it with \textit{Random Service Allocation} (RSA), Greedy service allocation and optimal solution derived by \textit{brute-force search} of Eq.1 for single applications and Eq.2 for group-based applications. We used the following metrics to measure throughput of the service allocation algorithms:

\small
$$ MuSIC_{Throughput}=\frac{MuSIC~output}{Optimal~Solution~of~Eq.~1} \times 100$$
$$ RSA_{Throughput}=\frac{RSA~output}{Optimal~Solution~of~Eq.~1} \times 100$$
$$ Greedy_{Throughput}=\frac{G-MuSIC~output}{Optimal~Solution~of~Eq.~1} \times 100$$
$$ G-MuSIC_{Throughput}=\frac{G-MuSIC~output}{Optimal~Solution~of~Eq.~2} \times 100$$
\normalsize

In RSA algorithm required services are randomly selected. In Greedy based algorithm, available services with maximum \emph{\textbf{total normalized QoS}} will be selected.

Fig. \ref{fig:MUSICRSA_Optimality} (a), (b) and (c) show the throughput of MuSIC, RSA and Greedy for mobile applications when there are several mobile users in the system with varied uncertainty. As it is shown in Fig. \ref{fig:MUSICRSA_Optimality} (a) MuSIC could achieve to 66\% performance when there are 100 mobile users. Fig. \ref{fig:MUSICRSA_Optimality} (c), (d) show the same results for RSA and Greedy algorithms. RSA have around 48\% performance when there are 100 mobile users in the systems. This unchanged performance in RSA throughput makes sense while RSA randomly assigns services to mobile users without using user trajectory information. Greedy algorithm could reach at the best about 60\% performance.

Fig. \ref{fig:MUSICRSA_Optimality} (d) shows the performance of G-MuSIC with different number of equal-sized groups for 100 users. As it can be understood from Fig. when there are 4 groups (each group has 25 members) the performance is about 52\%. This will increase to 65\% when there 20 groups (4 member each). Theses results make sense while with smaller group size, the average distance of users to center of group mobility is less than the larger size group. This makes MuSIC to find closer services to mobile users with higher QoS.

\begin{figure}
\begin{center}
\includegraphics[width=4.35in,height=4.35in]{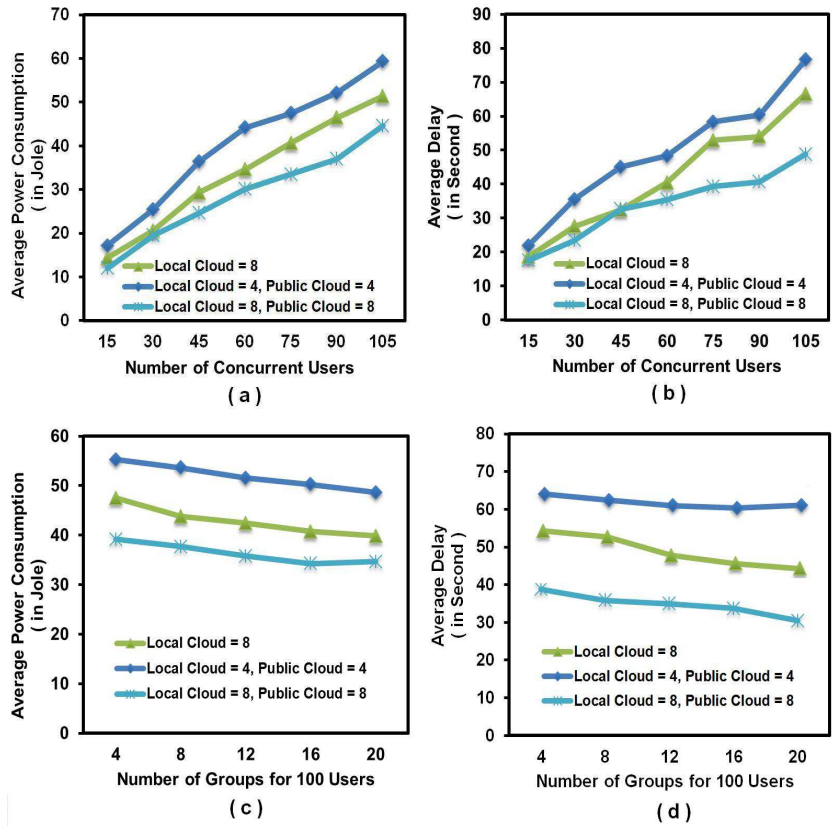}
\caption{MuSIC and G-MuSIC algorithm real averaged values for delay and power consumption}
\label{fig:MUSIC_Real_Data}
\end{center}
\vspace{-0.1in}
\end{figure}

Fig. \ref{fig:MUSIC_Real_Data} (a) and (b) show the real delay and power consumption according for different number of mobile users with LTW uncertainty in the range of [0\%,30\%]. For example as shown in Fig. \ref{fig:MUSIC_Real_Data} (a), by having 8 local clouds the average power consumption would be 50 jole/person (when there are 100 users). Adding 8 public services could decrease the power consumption about 20\%. This shows that increasing computing and storage resources does not necessarily increase the performance linearly while the communication bandwidth is a bottleneck.

\vspace{0.07in}
\textsl{\textbf{2-Tier Cloud Architecture Performance Study :}}
\vspace{0.07in}

In this section we study the performance of the 2-tier cloud architecture in comparison to only using public cloud services. One way of comparing the 2-tier cloud architecture is using the metrics described in table \ref{tb:metrics}. For example if a mobile application should experience constant low delay such as in some video streaming, then we could measure the gain that we get in power consumption and price by using 2-tier cloud architecture in comparison to only using public cloud. For example by using 2-tier cloud services the average power consumed on user device is 8 joules. If only public cloud is used then it will be 10 joules (due to long delay). Then the mobile user will gain $[1-8/10]\times100=20\%$ by using this 2-tier architecture in comparison to only using public cloud. By averaging this metric over all of the mobile users $u_k$ we could gain the average mobile users gain. The same procedure could be extended to power and price according to the table (\emph{Amazon large Instance} \cite{Amazon_WS} and \emph{T-Mobile} \cite{TMobile} prices were used as the data and cloud price model).

\begin{table}[b]
\begin{center}
\includegraphics[width=3.8in,height=2.8in]{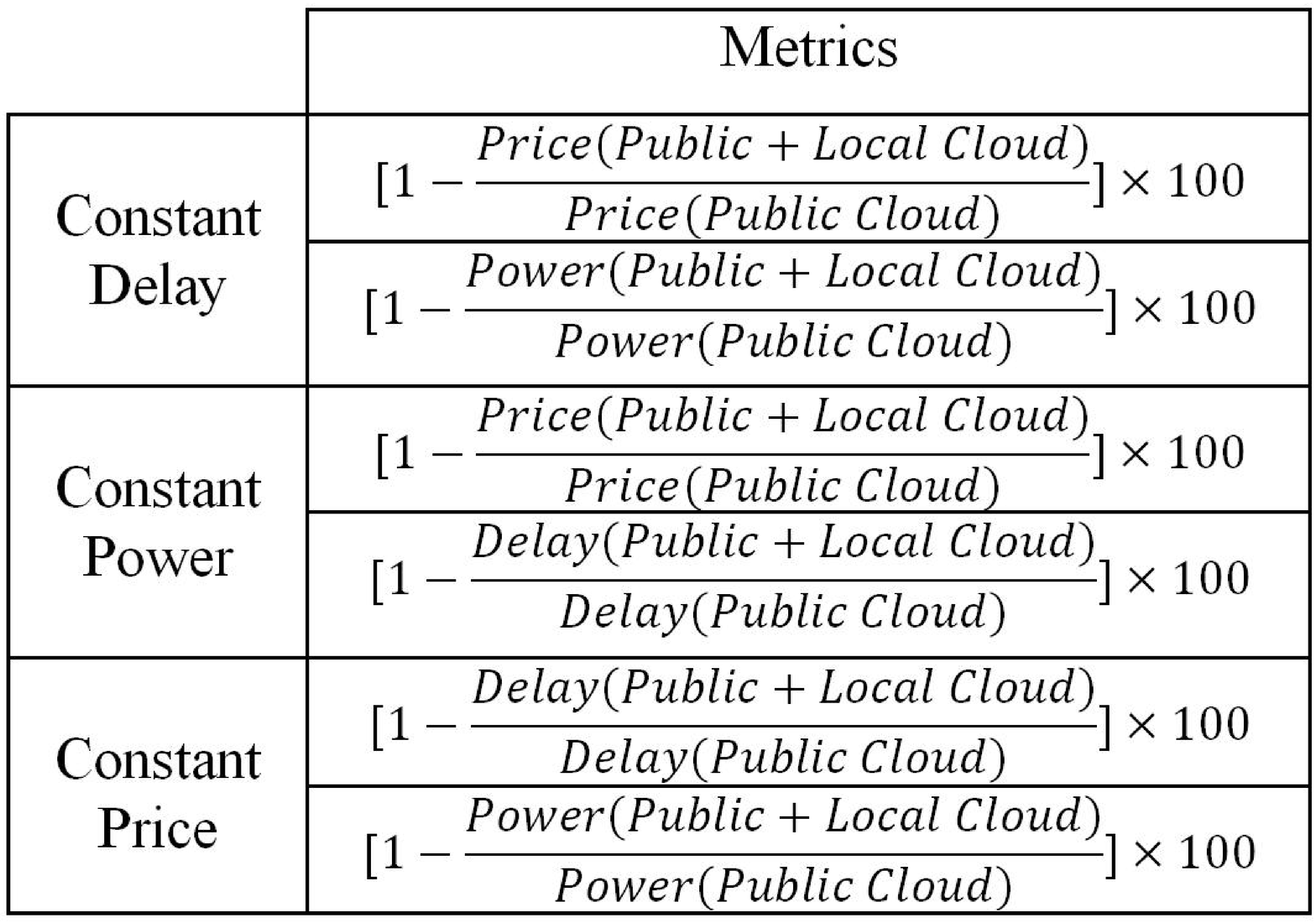}
\caption{ Performance Metrics for Evaluation of 2-Tier Cloud Architecture }
\label{tb:metrics}
\end{center}
\vspace{-.1in}
\end{table}

Table. \ref{tb:2tierperformance} shows the values of the mentioned metrics for 100 users using MuSIC, G-Music, Greedy and RSA algorithms with varied uncertainty in location-time prediction (in G-MuSIC we averaged over different groups). Having [0\%,30\%] uncertainty, with constant delay by using MuSIC one could get 27\% gain in price and 2\% gain in power consumption. These results would be 12\% and 2\% for G-MuSIC. They are intuitively correct while the lower the delay the lower power consumption is. The same is true for price because of cheaper price of WiFi and local services in comparison to 3G and public clouds virtualized services.
With constant power consumption by using MuSIC and G-MusIC one could get 22\% and 9\% percent saving in price and 4\% and 3\% percents in delay. With the constant price one could get 17\% decrease in power consumption and 15\% decrease in application delay using MuSIC. These results would be 8\% and 3\% for G-MuSIC.

As it could be understood from table MuSIC performance is better than using RSA and Greedy based approach. Having [0\%,30\%] uncertainty, with constant delay by using Greedy one could get 18\% gain in price and 5\% gain in power consumption. These results would be 10\% and 3\% for RSA. With constant power consumption by using Greedy and RSA one could get 14\% and 13\% percent saving in price and 2\% and 2\% percents in delay. With the constant price one could get 10\% decrease in power consumption and 12\% decrease in application delay using Greedy approach. These results would be 7\% and 10\% for RSA.

\begin{table*}
\begin{center}
\includegraphics[width=4.85in,height=2in]{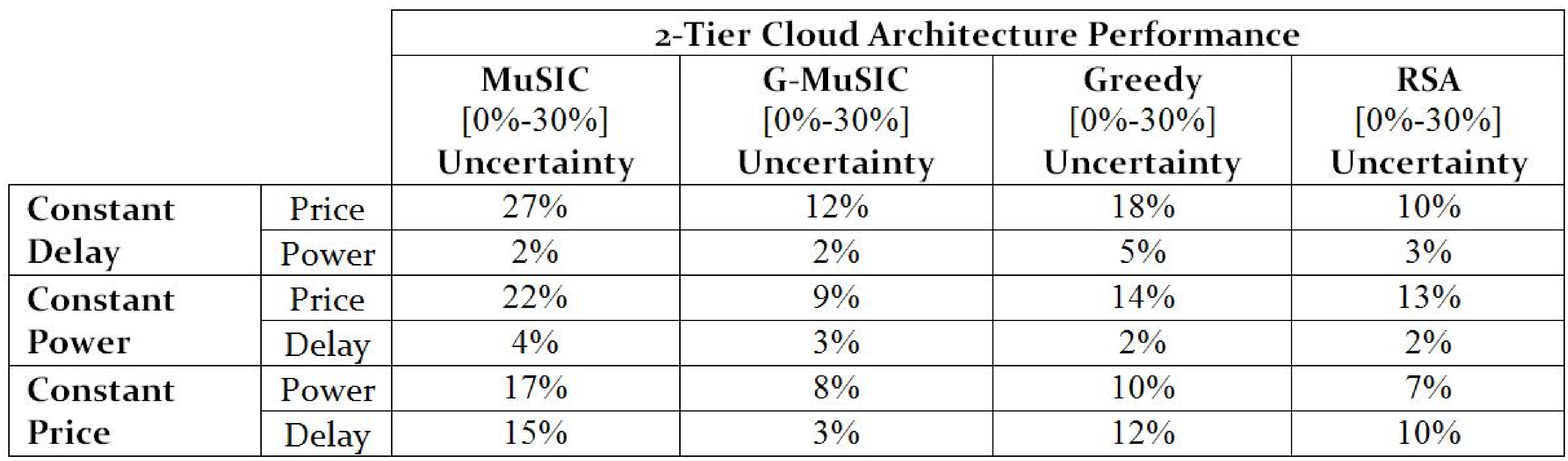}
\caption{ 2-Tier Cloud Performance Results}
\label{tb:2tierperformance}
\end{center}
\vspace{-.1in}
\end{table*}

\section{Related Works}\label{relatedwork}
The idea of remote execution of resource-intensive tasks to alleviate resource constraints in mobile device is not new in itself. The typical application runs a simple GUI on the mobile device and intensive-processing tasks on a remote server, \cite{Balan_2007}, \cite{Ou_2007}, \cite{Kumar_2010}, \cite{Flinn_2002}. Efficient execution of mobile applications by leveraging  \emph{grid computing platforms} has been addressed in systems such as \emph{MAPGrid} \cite{Huang_2007}. In \emph{MapGrid} \cite{Huang_2007}, intermittently available resources on grid platforms have been used to intelligently process and cache data for rich mobile applications such as video streaming. However, adapting the above techniques to work in the current cloud framework brings in new challenges and constraints. The autonomy of cloud resources leads to challenges in using the cloud effectively for mobile applications. In a grid environment, a grid proxy can provide storage, computational and network resources and it is often enough to find one resource node to service a mobile request. However, in the cloud environment, e.g. Amazon cloud services, storage and computational resources may be provided independently and charged individually. A single resource discovery process (for a request) may now need to be partitioned into multiple requests, one for each type of resource. The fact that users have to pay for public cloud resources also impacts the utility of these resources in the overall framework.
The \emph{Cloudlets} \cite{Satyanarayanan_2009} platform provides mechanisms for creation of resources near access points (AP) that provide computational and storage services for mobile users. Other efforts \cite{Giurgiu_2009}, \cite{MAUI_2010}, \cite{CloneCloud_2011}, use concepts from workflow technologies to partition applications between the mobile device and a local cloud. In particular, parameters such as code-size, allocated memory and computational needs of the application are shown to be crucial in effective partitioning of the workflow for high utility \cite{Giurgiu_2009}. The \emph{MAUI} \cite{MAUI_2010}, \emph{CloneCloud} \cite{CloneCloud_2011} and \cite{tocloud_2012} systems enable fine-grained energy-aware offloading of mobile application to the infrastructure. In particular, \emph{CloneCloud} uses static and dynamic application profilers to optimize execution of mobile applications in terms of energy consumption. Mechanisms to offload the execution tasks include method shipping (in \emph{MAUI} ) and on-demand delivery of execution state to pre-instantiated threads.

\emph{Spectra} \cite{Spectra_2002}, \emph{Chroma} \cite{Chroma_2003} and \emph{Cuckoo} \cite{Cuckoo_2010} are systems that use client-server architecture for offloading resource intensive tasks. In those systems the RPC is used to invoke the functionality from the server. In Spectra there is a registry which contains information about Spectra available servers, CPU loads, etc. Programmers need to manually partition the application by specifying which methods might be offloaded. In Spectra energy consumption and performance are considered as the criteria for task offloading. Spectra monitors constantly the resources such as CPU, network and battery to find the best service partitioning strategy. In Chroma an approach called "\emph{tactics}" is used. The system history is logged and machine learning techniques are used to do optimization for resource usage.
Cuckoo can offload tasks onto any resource that runs the Java Virtual machine, like public and local clouds. Cuckoo's main objectives are to enhance performance and reduce battery usage. In Cuckoo the application should be written in a way that supports remote execution as well as local execution. It uses the current Android programming model \emph{activity/service}. The services are candidates for offloading and activities are candidates that could be done locally. There are some gaps that should be filled out in this work like considering the mobility issues on system performance and price of the services on different cloud type like public cloud and local cloud.

There are some other approaches based on parallel processing of mobile applications such as \emph{Hyrax} and \cite{Canepa_2010}. In \emph{Hyrax} \cite{Hyrax_2010}, a system architecture based on \emph{MapReduce} \cite{mapreduce_2008} architecture has been proposed. Hyrax offloads intensive data and computational tasks on mobile platforms. Like Hadoop it has four main elements: one \emph{NameNode}, one \emph{JobTracker}, several \emph{DataNodes}, and \emph{TaskTrackers}. Jobs are scheduled and coordinated by NameNode and JobTracker among TaskTrackers. DataNodes store and provide access to data while TaskTrackers execute tasks assigned to them by JobTracker. The central server doesn't do anything about processing just scheduling the job among mobile devices. In Hyrax phones communicate with each other using WiFi. It inherits the fault tolerant property from Hadoop which recovers from task failure by re-execution and redundancy. Although it has a nice and scalable architecture, the performance of Hyrax is poor for CPU-bound tasks.

\emph{SCAMPI} \cite{sigcomm_Scampi_2012} supports distributed task execution in opportunistic pervasive network environment. It uses the human social behavior as the key element for optimal allocation for variety of services like sensors, personal communication devices and resources embedded in the local environment. SCAMPI borrows modeling framework from SOC as the abstraction for different services modeling. There are some gaps that should be considered like effect of users mobility patterns on system utility, different QoS criteria like power and price of each service for optimal service allocation.

MobiCloud \cite{MobiCloud_2010} proposed to use cloud computing to empower \emph{MANETs} (mobile ad hoc networks) in a secure way. In MobiCloud MANETs is transferred into service oriented architecture. Each node is considered as a \emph{Service Node} that can be used as a service provider or a service broker based on its computation and communication capabilities. Each service node is incorporated and mirrored on to the cloud as a virtualized component. These \emph{Extended Semi-Shadow Images} (ESSIs) are not exactly the same as \emph{virtual images} since an ESSI could be an \emph{exact clone}, a \emph{partial clone}, or merely an image that has \emph{extended functions} of the physical device. By using these ESSIs a virtualized routing and communication layer is established to assist the physical mobile nodes that they represent. MobiCloud does not present any experimental results to show the performance of the mobile applications and could be considered as the future vision of MCC resource allocation.

In \cite{Canepa_2010}, they proposed the  architecture based on group of mobile devices to upload the task. They claimed that this architecture could improve the mobile application performance but they did not considered the performance of the application such as power and delay which are critical for mobile applications.

In \emph{WhereStore} \cite{Stuedi_2010}, the authors considered the data sharing application. They showed that the locality of these storage can significantly improve the performance of the application, specially for location-based data search and sharing. In this work they mainly target to reduce the missing rate of replicas in such applications.

\section{Conclusions and Future Directions} \label{Conclusion}
In this paper we proposed a new  framework to model mobile applications as a Location-Time Workflow - the unique aspect here is that this abstraction models the mobile user service usage patterns based on user mobility. Our main goal was to  use this concept to optimally decompose the set of tasks to execute on the mobile clients and the 2-tier cloud architecture for two different type of mobile applications one is single user and the other one is collaborative mobile applications. We proposed an efficient algorithm called \textsl{MuSIC} that is able to achieve about 78\% of optimal solutions when the number of mobile users is high. Our studies also show that MuSIC performs quite well under uncertainty in prediction of mobile user LTW and different mobility patterns like random waypoint and Manhattan models. In our future work we will focus to make MAPCloud optimal for mobile games.



\begin{thebibliography}{}
\bibitem{Music_2013} M. Reza. Rahimi, Nalini Venkatasubramanian, Athanasios Vasilakos, "\emph{MuSIC: On Mobility-Aware Optimal Service Allocation in Mobile Cloud Computing}", In the IEEE 6th International Conference on Cloud Computing, (Cloud 2013), July 2013, Silicon Valley, CA, USA.
\bibitem{Sokol_2013} Marco Valerio, Sokol Kosta,Alessandro Mei,Julinda Stefa: " T\emph{o Offload or Not to offload? The Bandwidth and Energy Costs for Mobile Cloud Computing}", In IEEE INFOCOM 2013.
\bibitem{MAPCloud_2012} M. Reza. Rahimi, Nalini Venkatasubramanian, Sharad Mehrotra and Athanasios Vasilakos, "\emph{MAPCloud: Mobile Applications on an Elastic and Scalable 2-Tier Cloud Architecture}", In the 5th IEEE/ACM International Conference on Utility and Cloud Computing (UCC 2012), Nov 2012, USA.
\bibitem{RezaWoWMoM_2012} M. Reza. Rahimi, Nalini Venkatasubramania "\emph{Exploiting an Elastic 2-Tiered Cloud Architecture for Rich Mobile Applications}", In the IEEE/ACM 13th International Symposium on a World of Wireless, Mobile and Multimedia Networks (WoWMoM 2012), June 2012, USA.
\bibitem{sigcomm_Scampi_2012} Mikko Pitkänen, Teemu Kärkkäinen, Jörg Ott, Marco Conti, Andrea Passarella, Silvia Giordano, Daniele Puccinelli, Franck Legendre, Sacha Trifunovic, Karin Hummel, Martin May, Nidhi Hegde, and Thrasyvoulos Spyropoulos, "\emph{SCAMPI: service platform for social aware mobile and pervasive computing}", In Proceedings of the first edition of the MCC workshop on Mobile cloud computing (MCC '12, SIGCOMM'12), New York, NY, USA.
\bibitem{cloudstorm_2012}Sharon Choy, Bernard Wong, Gwendal Simon, and Catherine Rosenberg, "\emph{The Brewing Storm in Cloud Gaming: A Measurement Study on Cloud to End-User Latency}", In the Proceedings of the Workshop on Network and Systems Support for Games (NetGames), Venice, Italy, November 2012.
\bibitem{tocloud_2012} Vinod Namboodiri, Toolika Ghose "\emph{To Cloud or Not to Cloud: A Mobile Device Perspective on Energy Consumption of Applications}", In the IEEE/ACM 13th International Symposium on a World of Wireless, Mobile and Multimedia Networks (WoWMoM 2012), June 2012, USA.
\bibitem{RezaCCgrid_2012} M. Reza. Rahimi, Nalini Venkatasubramania "\emph{Cloud Based Framework for Rich Content Mobile Applications}", Poster In the IEEE/ACM 11th International Symposium on Cluster, Cloud and Grid Computing (CCGRID2011), Newport Beach, May 2011, USA
\bibitem{PARM_2011} Shivajit Mohapatra, M. Reza. Rahimi, Nalini Venkatasubranian "Power-Aware Middleware for Mobile Applications", Chapter 10 of the Handbook of Energy-Aware and Green Computing, Chapman \& Hall/CRC, 2011.
\bibitem{CloneCloud_2011} Byung-Gon Chun, Sunghwan Ihm, Petros Maniatis, Mayur Naik, Ashwin Patti, " \emph{CloneCloud: Elastic Execution between Mobile Device and Cloud}", In ACM European Conference on Computer Systems,  EUROSYS 2011.
\bibitem{cloudsim_2011} Rodrigo N. Calheiros, Rajiv Ranjan, Anton Beloglazov, Cesar A. F. De Rose, and Rajkumar Buyya, " \emph{CloudSim: A Toolkit for Modeling and Simulation of Cloud Computing Environments and Evaluation of Resource Provisioning Algorithms}", Wiley Press, 2011.
\bibitem{3gwifi_2010} Aruna Balasubramanian, Ratul Mahajan, Arun Venkataramani " \emph{Augmenting Mobile 3G Using WiFi: Measurement, Design, and Implementation}" In the Proceedings of the International Conference on Mobile Systems, Applications, and Services, \emph{MobiSys 2010}.
\bibitem{power_tutor_2010} Lide Zhang, Birjodh Tiwana, Zhiyun Qian, Zhaoguang Wang, Robert P. Dick, Zhuoqing Morley Mao, and Lei Yang " \emph{Accurate online power estimation and automatic battery behavior based power model generation for smartphones}" In the Proceedings IEEE/ACM/IFIP CODES/ISSS, 2010.
\bibitem{Canepa_2010} G. H. Canepa, D. Lee " \emph{A Virtual Cloud Computing Provider for  Mobile Devices}", In Proceedings of the 1st ACM Workshop on Mobile Cloud Computing;  Services: Social Networks and Beyond (\emph{MCS '10}), New York, 2010.
\bibitem{Kumar_2010} Karthik Kumar, Yung-Hsiang Lu " \emph{Cloud Computing for Mobile Users: Can Offloading Computation Save Energy?}", \emph{Computer}, Apr. 2010.
\bibitem{above_thecloud} M. Armbrust, A. Fox, R. Griffith, Anthony D. Joseph, R. Katz, A. Konwinski, G. Lee, D. Patterson, A. Rabkin, I. Stoica, and M. Zaharia, " \emph{A View of Cloud Computing} " Commun. ACM 53, 4 (April 2010).
\bibitem{MAUI_2010} Eduardo Cuervo, Aruna Balasubramanian, Dae-ki Cho, Alec Wolman, Stefan Saroiu, Ranveer Chandra, and Paramvir Bahl " \emph{MAUI: Making Smartphones Last Longer with Code Offload}", In The International Conference on
    Mobile Systems, Applications, and Services, \emph{MobiSys 2010}.
\bibitem{DB_2010} Avi Silberschatz, Henry F. Korth, S. Sudarshan, "\emph{Database System Concepts}", Sixth Edition, McGraw-Hill, 2010.
\bibitem{Hyrax_2010} E. Marinelli " \emph{Hyrax: Cloud Computing on Mobile Device using MapReduce}", Master Thesis Draft, Computer Science Department, CMU, Sept. 2009.
\bibitem{Juniper_2010} Juniper, "\emph{Mobile Cloud Computing: \$9.5 billion by 2014}", Juniper, http://www.readwriteweb.com/archives, Tech. Rep., 2010.
\bibitem{Stuedi_2010} P. Stuedi, I. Mohomed, and D. Terry " \emph{WhereStore: location-based data storage for mobile devices interacting with the cloud}", In the Proceedings of the 1st ACM Workshop on Mobile Cloud Computing \& Services, \emph{MCS '10} New York, NY, USA, 2010.
\bibitem{Cuckoo_2010} R. Kemp, N. Palmer, T. Kielmann, and H. Bal "\emph{Cuckoo: A Computation offloading Framework for Smartphones}", In IEEE International Conference on Mobile Computing, Applications and Services, \emph{MOBICASE 2010}.
\bibitem{MobiCloud_2010} Dijiang Huang, Xinwen Zhang, Myong Kang, and Jim Luo, "\emph{MobiCloud: Building Secure Cloud Framework for Mobile Computing and Communication}", In Proceedings of the 2010 Fifth IEEE International Symposium on Service Oriented System Engineering, \emph{SOSE 2010}.
\bibitem{Satyanarayanan_2009} M. Satyanarayanan, P. Bahl, R. Cáceres, N. Davies " \emph{The Case for VM-Based Cloudlets in Mobile Computing}", In the IEEE International Conference on Pervasive Computing and Communication (PerCom) (\emph{PerCom}), 2009.
\bibitem{Giurgiu_2009} I. Giurgiu, O. Riva, D. Juric, I. Krivulev and G. Alonso " \emph{Calling The Cloud: Enabling Mobile Phones as Interfaces to Cloud Applications}",  In the Proceedings of the 10th ACM/IFIP/USENIX international Conference on \emph{Middleware2009}.
\bibitem{Mabrouk_2009} N. B. Mabrouk, S. Beauche, E. Kuznetsova, N. Georgantas, and V. Issarny " \emph{Qos-aware Service Composition in Dynamic Service Oriented Environments}", In the Proceedings of the 10th ACM/IFIP/USENIX international Conference on \emph{Middleware2009}.
\bibitem{Qos_2009} Nebil Ben Mabrouk, Nikolaos Georgantas, and Valerie Issarny, "\emph{A semantic end-to-end QoS model for dynamic service oriented environments}", In Proceedings of the 2009 ICSE Workshop on Principles of Engineering Service Oriented Systems (PESOS '09), Washington, DC, USA,
\bibitem{ABI_2009} ABI, "\emph{Mobile Cloud Computing Subscribers to Total Nearly one Billion by 2014}", ABI http://www.abiresearch.com/press/1484/, Tech. Rep., 2009.
\bibitem{Vaquero_2008} L. M. Vaquero, L. Rodero-Merino, J. Caceres, and M. Lindner " \emph{A Break in The Clouds: Towards a Cloud Definition}", In ACM Special Interest Group on Data Communication,  \emph{SIGCOMM} Computer. Communication. Rev. 39, 1 (Dec. 2008).
\bibitem{mapreduce_2008} Jeffrey Dean, Sanjay Ghemawat " \emph{MapReduce: simplified data processing on large clusters}", Commun. ACM 51, 1 (January 2008), 107-113.
\bibitem{Balan_2007} R. K. Balan, D. Gergle, M. Satyanarayanan, and J. Herbsleb " \emph{Simplifying cyber foraging for mobile devices}", In \emph{MobiSys}, 2007.
\bibitem{Huang_2007} Y. Huang, N. Venkatasubramanian " \emph{MAPGrid: A New Architecture for  Empowering Mobile Data Placement in Grid Environments}", In Proceedings of IEEE/ACM International Symposium on Cluster, Cloud and Grid Computing, CCGrid 2007.
\bibitem{Ou_2007} S. Ou, K. Yang, and J. Zhang "\emph{An Effective offloading Middleware for Pervasive Services on Mobile Devices}", Journal of Pervasive and Mobile Computing, 2007.
\bibitem{Zeng_2004} L. Zeng, B. Benatallah, A. H. NGU, M. Dumas, J. Kalagnanam, and H. Chang " \emph{Qos-Aware Middleware for Web Services Composition }", In \emph{IEEE Trans. Software. Eng}, 2004.
\bibitem{RWP_2004} Christian Bettstetter, Hannes Hartenstein, and Xavier Prez-Costa, "\emph{Stochastic properties of the random waypoint mobility model}", Wirel. Netw. 10, 5 (September 2004).
\bibitem{Shenoy_2003} P. Shenoy and P. Radkov " \emph{Proxy-assisted Power-friendly Streaming  to Mobile Devices}", In Proceedings of the Multimedia Computing and Networking (\emph{MMCN})  Conference, 2003.
\bibitem{Flinn_2002} Jason Flinn, SoYoung Park, and M. Satyanarayanan " \emph{Balancing Performance, Energy, and Quality in Pervasive Computing }", In Proceedings of the 22 ndInternational Conference on Distributed Computing Systems (\emph{ICDCS'02}).
\bibitem{simulated_anealing} R. W.  Eglese, " \emph{Simulated Annealing: A tool for Operational Research}", European Journal of Operational Research, 1990.
\bibitem{Chroma_2003}R. Balan, M. Satyanarayanan, S. Park, T. Okoshi, "\emph{Tactics-based Remote Execution for Mobile Computing}", In \emph{MobiSys 2003}.
\bibitem{Spectra_2002} J. Flinn, S. Park, M. Satyanarayanan, "\emph{Balancing Performance, Energy, and Quality in Pervasive Computing}", In Proceedings of the IEEE International Conference on Distributed Computing Systems, \emph{ICDCS 2002}.
\bibitem{Mobility_2002} T. Camp, J. Boleng, and V. Davies, "\emph{A Survey of Mobility Models for Ad Hoc Network Research, in Wireless Communication and Mobile Computing (WCMC)}", Special issue on Mobile Ad Hoc Networking: Research, Trends and Applications, vol. 2, no. 5, pp. 483-502, 2002.
\bibitem{Knapsack_1998} A. Drexl, "\emph{A Simulated Annealing Approach to the Multiconstraint Zero-One Knapsack Problem}", Springer Computing 1988, Volume 40, Issue 1, pp 1-8.
\bibitem{Amazon_WS} Amazon Web Services: http://aws.amazon.com/
\bibitem{Wowza} Wowza Media Server 2 for Amazon EC2 http://www.wowza.com/
\bibitem{Azure} Microsoft Windows Azure: http://www.microsoft.com/windowsazure/
\bibitem{Google_App_Engine} Google Application Engine: http://code.google.com/appengine/
\bibitem{TMobile} T-Mobile Data Plan: http://www.t-mobile.com/shop/plans/
\bibitem{MYSQL} MySQL DB http://www.mysql.com/

\end{thebibliography}
\end{document}